\documentclass[10pt,twocolumn,twoside,compsoc]{IEEEtran}

\usepackage{graphicx}
\usepackage{caption}
\usepackage{subfig}
\usepackage{cite}
\usepackage{stmaryrd}
\usepackage{amsfonts,latexsym,amssymb}
\usepackage[cmex10]{amsmath}
\usepackage{algorithm}
\usepackage{algorithmic}
\usepackage{multirow}
\usepackage[table,svgnames]{xcolor}

\usepackage{hyperref}
\input{zgx.math}

\newcommand{\figbkcolor}{AliceBlue}

\begin{document}
\bstctlcite{IEEEtran:BSTcontrol}

\title{Linked Component Analysis \\
from Matrices to High Order Tensors:  \\ Applications to Biomedical Data}
\author{Guoxu~Zhou, Qibin~Zhao, Yu Zhang,  T\"ulay Adal\i, Shengli Xie, and Andrzej~Cichocki 
\IEEEcompsocitemizethanks{\IEEEcompsocthanksitem Guoxu Zhou is with the Laboratory for Advanced Brain Signal Processing, RIKEN, Brain Science Institute, Wako-shi, Saitama, Japan and  the School of Automation at Guangdong University of Technology, Guangzhou, China (Corresponding author, e-mail: zhouguoxu@brain.riken.jp).}
\IEEEcompsocitemizethanks{\IEEEcompsocthanksitem Qibin Zhao is with the Laboratory for Advanced Brain Signal Processing, RIKEN, Brain Science Institute, Wako-shi, Saitama 3510198, Japan (e-mail: qbzhao@brain.riken.jp).}
\IEEEcompsocitemizethanks{\IEEEcompsocthanksitem Yu Zhang is with the Key Laboratory for Advanced Control and Optimization for Chemical Processes, Ministry of Education, East China University of Science and Technology, Shanghai 200237, China (e-mail: yuzhang@ecust.edu.cn)).}
\IEEEcompsocitemizethanks{\IEEEcompsocthanksitem T\"ulay Adal{\i} is with the Department of Computer Science and Electrical Engineering, University of Maryland, Baltimore County, MD 21250 USA (e-mail: adali@umbc.edu).}
\IEEEcompsocitemizethanks{\IEEEcompsocthanksitem Shengli Xie is with the School of Automation at Guangdong University of Technology, Guangzhou, China (e-mail: shlxie@gdut.edu.cn).}
\IEEEcompsocitemizethanks{\IEEEcompsocthanksitem Andrzej Cichocki is with the Laboratory for Advanced Brain Signal Processing, RIKEN, Brain Science Institute, Wako-shi, Saitama 3510198, Japan, and with Systems Research Institute, Polish Academy of Science, Warsaw, Poland (Corresponding author, e-mail: cia@brain.riken.jp).}

}

\IEEEaftertitletext{\vspace{-2.5\baselineskip}}

\markboth{Proceedings of THE IEEE}%
{ZHOU \MakeLowercase{\textit{et al.}}: Linked Component Analysis and its Applications to Biomedical Data }

\maketitle

\begin{abstract}
With the increasing availability of various sensor technologies, we now have access to large amounts of multi-block  (also called multi-set, multi-relational, or multi-view) data that need to be jointly analyzed to explore their  latent connections. Various component analysis methods have played an increasingly  important role for the analysis of such coupled data. In this paper, we first provide a  brief review of existing matrix-based (two-way) component  analysis methods for the joint analysis of such data with a focus on biomedical applications. Then, we discuss their important extensions and generalization to multi-block multiway (tensor) data. We show how constrained multi-block tensor decomposition methods are able to extract similar or statistically dependent common features that are shared by all blocks, by incorporating the multiway nature of data. Special emphasis is given to the flexible common and individual feature analysis of multi-block data with the aim to simultaneously extract common and individual latent components with desired properties and types of diversity. Illustrative  examples are given to demonstrate their effectiveness for biomedical data analysis.
\end{abstract}


\section{Introduction}

Development and wide deployment of various  sensor technologies have resulted in the generation of large amounts of multi-block data that need to be jointly analyzed in order to extract physiologically  meaningful and useful  hidden (latent) components. Medical recording and diagnostic devices play an important role among those.
In order to explore spatial, temporal, and spectral differences and similarities in multi-array, multi-block  bio-signals and images, researchers developed a variety of  strategies based on component analysis and  multiway factor analysis.
In this paper, we focus on analysis of   multi-block data that are naturally connected using tensor/matrix factorizations.
We  critically review  emerging multiway component analysis  or multilinear blind source separation approaches that  include procedures for extracting hidden components (or source signals) with specific features or constraints (e.g., sparsity, nonnegativity, statistical independence).  The presented approaches  can be applied not only to unsupervised multilinear blind source separation but also for missing data imputation (tensor completion), feature extraction, classification, clustering and anomaly detection, especially in application to electroencephalography (EEG), electrocorticography (ECoG), and magnetic resonance imaging (MRI) data.  Multi-block  tensor  decompositions and multi-way analysis allow us to discover meaningful hidden spatio-temporal structures of complex data and perform generalizations by capturing multi-linear and multi-aspect relationships. The challenge is how to analyze   complex inhomogeneous multi-set large-scale multi-array data.  We describe the key procedure to apply tensor/matrix  factorization/decompositions methods  for these problems in practice and demonstrate a number of successful examples.

Tensors have been adopted in  diverse branches of data analysis, such as in signal and image processing, psychometrics, chemometrics, biometrics, quantum physics, quantum chemistry and neuroscience \cite{NMF-book,Kolda09tensordecompositions,SPM_Tensor, SCIECIA, Hackbush2012,Smilde,Kroonenberg,acar2013understanding,acar2012coupled,jafari2005fetal}.
 Tensor decompositions provide extensions of  blind source separation (BSS) and 2-way (matrix) component analysis (2-way CA) to  multi-way component analysis (MWCA) \cite{SPM_Tensor, SCIECIA}, and have proven to be suitable for dimensionality reduction of multi-way data, even if the data are incomplete and corrupted by noise.
Tensor decompositions are particularly attractive for biomedical data analysis that exhibits not only large volume but also very high variety. Thus, they are suited to problems in bio- and neuro-informatics or computational  neuroscience  where data  are collected in various forms of large, sparse tabular,  graphs or networks  with multiple aspects and high dimensionality. For historical  overview of tensor factorizations/decompositions and important  scientific contributions 
in these areas  in some chronological orders please check \cite{Kolda09tensordecompositions,SPM_Tensor} and references therein.

Moreover, multi-block  tensors, which arise in numerous important applications that require the
analysis of  diverse and partially related data, can be decomposed to common (or correlated) and individual (or uncorrelated)  components.
The effective analysis of coupled tensors requires the development of new models and associated
algorithms that can identify the core relations that may exist among the different tensors, and scale to large multi-sets data.

This  paper extends beyond the standard tensor decomposition models such as the unconstrained  single block Tucker and canonical polyadic (CP) models, and aims to demonstrate flexibilities  of matrix/tensor decompositions for multi-block and  multi-modal data, together with their role as a mathematical backbone for the discovery of hidden structures in  large-scale multi-relational data \cite{NMF-book,SPM_Tensor}.

\subsection{Preliminaries -- Basic Tensor Notations and Operations}

A higher-order tensor can be interpreted as a multiway array of numbers.  Tensors are denoted  by  calligraphic capital letters, e.g., $\tensor{X}  \in \Real^{I_{1} \times I_{2} \times \cdots \times I_{N}}$
 (we shall assume that all entries of a tensor are real-valued). The order  of a tensor is the number of its ``modes'', ``ways'' or ``dimensions'', which include e.g., space, time, frequency,  trials, subjects, classes, and dictionaries \cite{NMF-book,Kolda09tensordecompositions}. A matrix has two modes: rows and columns, while an $N$th-order tensor has $N$ modes.

Matrices (2nd-order tensors) are denoted by boldface capital letters, e.g., $\mat{X}$, and vectors (1st-order tensors)  by boldface lowercase letters; for instance  the columns of the matrix $\mat{A}=[\mat{a}_1,\mat{a}_2, \ldots,\mat{a}_R]  \in \Real^{I \times R}$ are denoted by $\mat{a}_r$ and elements of a
matrix (scalars) are denoted by lowercase letters, e.g., $a_{i,r}$ (or $[\mat{A}]_{i,r}$).

\begin{figure}[!t]
\centering
\includegraphics[width=.95\linewidth]{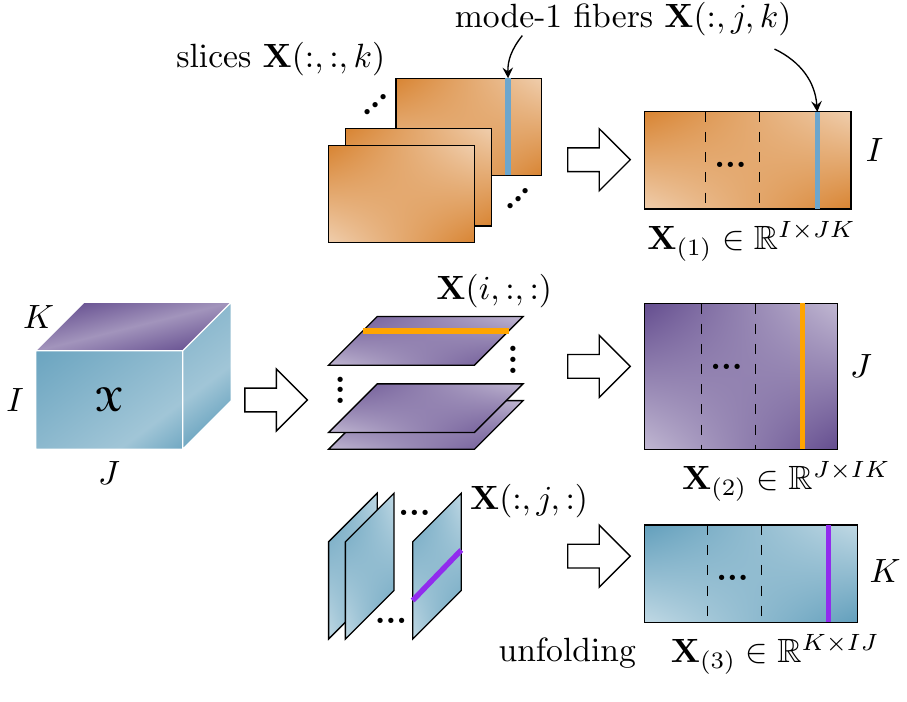}
\caption{Fibers, slices, and unfolding of a 3rd-order tensor.  }
\label{fig:tenNotations}
\end{figure}

Subtensors are formed when a subset of indices is fixed. Of particular interest are mode-$n$
 {\it fibers} (vectors),
 defined  by fixing every index but $n$,  and
{\it slices} which are two-dimensional sections (matrices) of a tensor,
obtained by fixing all the indices but two, see  \figurename \ref{fig:tenNotations}.

The process of unfolding  flattens a tensor into a  matrix \cite{Kolda09tensordecompositions}.
In the simplest scenario, mode-$n$ unfolding (matricization,  flattening) of  tensor
 $\tensor{X} \in \Real^{I_{1} \times I_{2} \times \cdots \times I_{N}}$ yields
a matrix $\tenmat{X} \in \Real^{I_{n} \times (I_{1} \cdots I_{n-1} I_{n+1} \cdots I_N)}$ whose columns are all the mode-$n$ fibers arranged in a specific order (in this paper rows and columns are ordered  colexicographically).
See \figurename \ref{fig:tenNotations} for an illustration.

The   mode-$n$  product of the tensor
$\tensor{X} \in \Real^{I_{1} \times \cdots \times I_{N}}$ and a matrix  $\mat{B} \in \Real^{J \times I_n}$
is the tensor
$ \tensor{C} = \tensor{X} \times_n \mat{B} \in \Real^{I_1 \times \cdots \times I_{n-1} \times J \times I_{n+1} \times \cdots \times I_N}$
with entries $ c_{i_1,i_2,\ldots,i_{n-1},j,i_{n+1},\ldots, i_{N}} =\sum_{i_n=1}^{I_n} x_{i_1,i_2,\ldots,i_N} \; b_{j,i_n}$.
This can be also expressed in a matrix form as $\mat{C}_{(n)} =\mat{B} \mat{X}_{(n)}$.

{The Kronecker product $\kkp$ and the Khatri-Rao product $\krp$ of matrices are defined in TABLE \ref{tab:notations}. Moreover, we define $\bigkkp_{p\neq n}\matn[p]{A}=\matn[N]{A}\kkp\cdots\kkp\matn[n+1]{A}\kkp\matn[n-1]{A}\cdots\kkp\matn[1]{A}$, and similarly define $\bigkrp_{p\neq n}\matn[p]{A}$. }

See Table \ref{tab:notations} for the other useful notations. We refer  readers to \cite{Kolda09tensordecompositions,SPM_Tensor} for more details about various tensor operations.

\newcommand{\dohang}{\hangindent2em\hangafter1 }
\newcolumntype{R}{%
  >{\everypar\expandafter{\the\everypar\dohang\everypar{\dohang}}\arraybackslash}%
  p{5.2in}%
}
 \begin{table*}[!t]
\caption{Notations And Definitions}
\label{tab:notations}
\centerline{
\colorbox{\figbkcolor}{
\begin{tabular}{  r   R }
\hline \hline
\mat{B}, \mats[r]{b},  $b_{m,r}$ & Matrix, the $r$th-column,  and the ($m,r$)th-entry of \mat{B}. $[\mat{B}]_{m,r}=b_{m,r}$.\\
\trans{\mat{B}}, $\trace{\mat{B}}$ & Transpose, trace of matrix \mat{B} \\
{$\krank{\mat{B}}$} & { Kruskal rank of \mat{B} is the maximum value ensuring that any subset of $\krank{\mat{B}}$ columns is linearly independent.} \\
$\tensor{X}$, \tenmat{X} & Tensor, the mode-$n$ matricization of tensor \tensor{X}. \\
{\frob{\mat{B}}} & { Frobenius norm of $\mat{B}\in\Real^{M\times R}$, i.e., $\frob{\mat{B}}=({\sum_{m,r}b_{mr}^2})^{\frac{1}{2}}$.  For tensor \tensor{X}, $\frob{\tensor{X}}=\frob{\tenmat{X}},\; \forall n$. } \\
$\hdp$, $\matdiv{}{}$  & Element-wise (Hadamard) product, division  of matrices or tensors. \\
$\mat{C}=\mat{A}\kkp\mat{B}$ &  Kronecker product of $\mat{A}\in\Real^{I_1\times J_1}$ and $\mat{B}\in\Real^{I_2\times J_2}$ yields  $\mat{C}=\left[a_{i_1j_1}\mat{B}\right]\in\Real^{I_1I_2\times J_1J_2}$ with entries $c_{(i_1-1)I_2+i_2,(j_1-1)J_2+j_2}=a_{i_1j_1}b_{i_2j_2}$.  \\
$\mat{C}=\mat{A}\krp\mat{B}$ 	& Khatri-Rao product of $\mat{A}=\begin{bmatrix}
\mats[1]{a} & \mats[2]{a} & \cdots & \mats[J]{a}
\end{bmatrix}\in\Real^{I_1\times J}$ and $\mat{B}=\begin{bmatrix}
\mats[1]{b} & \mats[2]{b} & \cdots & \mats[J]{b}
\end{bmatrix}\in\Real^{I_2\times J}$ yields a matrix $\mat{C}=\begin{bmatrix}
\mats[1]{a}\kkp\mats[1]{b} & \mats[2]{a}\kkp\mats[2]{b} & \cdots & \mats[J]{a}\kkp\mats[J]{b}
\end{bmatrix}\in\Real^{I_1I_2\times J}$.  \\
$\tensor{C}=\tensor{X}\times_n\mat{B}$ &  Mode-$n$  product of tensor
$\tensor{X} \in \Real^{I_{1} \times \cdots \times I_{N}}$ and matrix  $\mat{B} \in \Real^{J \times I_n}$
yields tensor
$ \tensor{C} = \tensor{X} \times_n \mat{B} \in \Real^{I_1 \times \cdots \times I_{n-1} \times J \times I_{n+1} \times \cdots \times I_N}$
with entries $ c_{i_1,i_2,\ldots,i_{n-1},j,i_{n+1},\ldots, i_{N}} =\sum_{i_n=1}^{I_n} x_{i_1,i_2,\ldots,i_N} \; b_{j,i_n}$ and matrix representation $\tenmat{C}=\mat{B}\tenmat{X}$.
\\
$\tensor{X}=\matn[1]{b}\outerp\matn[2]{b}\outerp\cdots\outerp\matn[N]{b}$ & Outer product of vectors $\matn{b}\in\Real^{I_n}$ yields a rank-1 tensor $\tensor{X}=\matn[1]{b}\outerp\matn[2]{b}\outerp\cdots\outerp\matn[N]{b}$ with entries $x_{i_1,i_2,\ldots,i_N} = b^{(1)}_{i_1} b^{(2)}_{i_2} \cdots b^{(N)}_{i_N}$. \\
\hline \hline
\end{tabular}}
}
\end{table*}

\section{Two-way  Blind Signal Separation}

Blind source separation (BSS) {is an unsupervised learning method with the aim of estimating} $R$ source signals 
$\mat{B}\in\Real^{T\times R}$ from the measurement matrix $\mat{X}\in\Real^{I \times T}$, 
such that  \cite{Cichocki2002, ComonBSS2010} 
\begin{equation}
\label{eq:MixingModel}
\mat{X} = \mat{A}\trans{\mat{B}}
= \sum_{r=1}^R \mat{a}_r \mat{b}^T_r = \sum_{r=1}^R \mat{a}_r  \circ \mat{b}_r,
\end{equation}
where {the outer product $\outerp$ is defined in TABLE \ref{tab:notations}}, $\mat{A} =[\mat{a}_1,\mat{a}_2,\ldots, \mat{a}_R] \in \Real^{I \times R}$
is the unknown mixing matrix (also known as the basis matrix or dictionary, depending on application) and $\mat{B} =[\mat{b}_1,\mat{b}_2,\ldots, \mat{b}_R] \in \Real^{T \times R}$  is the matrix of sources \mats[r]{b} (components or latent variables).
Although we explicitly assume that the columns of \mat{B} are the latent variables, please keep in mind that the roles of \mat{A} and \mat{B} can be exchanged if we consider $\trans{\mat{X}}$ in \eqref{eq:MixingModel}.

Without any constraint, the matrix factorization problem \eqref{eq:MixingModel} is in general highly undetermined as it actually has an infinite number of solutions. However, 
with  very little \apriori information, 
BSS enables the recovery of sources in the sense that
\begin{equation}
\label{eq:uniqueBSS}
\mat{\hat{B}}=\Psi(\mat{X})=\mat{B\Lambda P},
\end{equation}
where \mat{\hat{B}} is an estimate of \mat{B}, $\Psi$ denotes a suitable BSS algorithm, \mat{\Lambda} is a diagonal scaling matrix and \mat{P} is a permutation matrix, denoting the unavoidable scaling and permutation ambiguities, respectively \cite{ComonBSS2010}.  This feature is called essential uniqueness of BSS.   From a more general sense,  \eqref{eq:MixingModel} essentially provides a representation or transformation of the data matrix \mat{X} that preserves as much of the information as possible. Different types of information to be preserved lead to various BSS criteria and methodologies. Below we briefly introduce the most popular ones.

\emph{Principal component analysis (PCA)}. PCA transforms data matrix $\mat{X}$ into a set of uncorrelated variables, called the principal components using either eigenvalue or 
singular value decomposition (EVD/SVD). A common use of PCA is for dimensionality reduction
where only a set of principal components {is} retained to preserve the maximum variance in the data. In PCA, uniqueness is achieved by imposing orthogonality on the transform matrix.

\emph{Independent component analysis (ICA)}. 
If we rewrite the matrix product in \eqref{eq:MixingModel} using the random vector notation 
\begin{equation}
\label{eq:traBSS}
{\mat{x}}(t)=\mat{A}{\mat{s}}(t),
\quad t=1,2,\ldots, T
\end{equation}
thus interpreting each row of
$\mat{X}$ as $T$-dimensional realization of the entries of the random vector $\mat{x}(t)$ and   the rows of 
$\trans{\mat{B}}$ in \eqref{eq:MixingModel} as the $T$-dimensional realization of the entries of the random vector $\mat{s}(t)$,
 then we can pose
the problem as one that estimates a demixing matrix $\mat{W}$
such that the rows of $\mat{Wx}$ are as independent as possible.
{In other words, the {demixing} matrix \mat{W} is expected to be the pseudo-inverse of the mixing matrix \mat{A}, neglecting the scaling and permutation ambiguities.}
With this principle, {we need a criterion to measure the statistical independence}, and the most natural criterion is the maximization of independence through minimization of mutual information which can 
be shown to include others such as maximum likelihood and maximization of non-Gaussianity \cite{AdaliIVASPM,adaliTSP2008, ComonBSS2010}. Another approach is to explicitly 
{estimate} higher-order statistics such as cumulants to establish independence using joint
diagonalizations as in joint approximate diagonalization of eigenmatrices (JADE) 
\cite{Cardoso1998}. 
ICA can be viewed as a generalization of PCA in the sense that it requires to venture beyond second-order statistics to incorporate higher-order
statistical information \cite{Cardoso1998}. 
{ICA is attractive because by only using the natural assumption of  statistical independence, 
it enables an essentially unique decomposition without the need for any additional 
constraint. For example, when both non-Gaussianity and sample dependence are taken into account in
the formulation, any source, including those that are Gaussians can be identified subject to permutation and scaling ambiguities as long as the
Gaussian sources do not have proportional covariances \cite{AdaliIVASPM}. }

ICA has proven to be very useful for functional MRI (fMRI) data analysis, and can be performed in two different ways \cite{calhoun2012,STICA2006,Calhoun-Adali2001spatial}: namely spatial ICA that extracts unique independent spatial maps, and temporal ICA that extracts independent time courses by considering \trans{\mat{X}} in \eqref{eq:MixingModel}. Although they are essentially just two different ways for organizing the data, spatial ICA is more useful as the spatial independence assumption is well suited for the systematically non-overlapping nature of the spatial patterns for most cognitive activation paradigms \cite{AdaliIVASPM, groupICA2008guo}.

\emph{Sparse component analysis (SCA)}. Sparsity is a natural property of many real-word signals, for example, activities of our brain, eye movements, and heart beat among others.  SCA is a technique that allows us to recover latent source signals by exploiting the sparsity of source signals and/or the mixing matrix; and it is even able to separate dense signals, because some smooth dense signals often can be very sparse in a transform domain \cite{marvasti2012unified}. For example, EEG signals are much more sparse in the frequency domain than  in the time domain \cite{sanei2013eeg}. In this case, BSS can be performed in the transform domain using SCA and then the outputs  can be inversely transformed to construct the original source signals. SCA is particularly important because, by incorporating  compressed sensing \cite{spm_cs2008}, it allows the solution of underdetermined BSS problem, i.e., when there are fewer observations than sources ($I<R$) in \eqref{eq:MixingModel}. Before we can do this, of course, we need to know the mixing matrix \mat{A}, which is often estimated by clustering the observations using the sparsity assumption on the sources \cite{TNN-sMixEst}, or by blind identification methods utilizing the  statistical independence of sources \cite{ FOBIUM, FOBIUM2007}. 

\emph{Nonnegative matrix factorization (NMF)}.  In  power spectral analysis for EEG data, the latent frequency components are expected to be nonnegative and generally very sparse. In this scenario, NMF proves to be particularly useful as it imposes nonnegativity on both the mixing matrix and the source signals \cite{Lee1999}. 
Due to the nonnegativity constraints, the data matrix is represented as purely additive combinations of basic building blocks while without any subtractions, endowing NMF with the special ability of learning-parts  \cite{Lee1999}.
 While NMF achieved great success in machine learning, lack of uniqueness guarantee in {the} general case poses  difficulty when it is applied to BSS.  Fortunately, by imposing additional sparsity---even very weak sparsity---on the sources, NMF is able to provide unique factorization  \cite{GillisSUNMF, HuangNMF2013}. Typically, under the pure-source dominance condition \cite{nLCAIVM2010, TNN-MVCNMF}, that is, for each signal there exists at least one instant at which only this signal is active or strongly dominant, NMF can be a powerful tool to separate statistically dependent nonnegative sources  \cite{nLCAIVM2010, TNN-MVCNMF}. 
NMF based on the pure-source dominance assumption is called separable NMF, which is highly scalable and can be applied to discover the most {representative members} in a large database \cite{FastSepNMF}. 

If we further require each row of \mat{B} to be pure-source dominant, i.e., the whole matrix \mat{B} to be a sub-matrix of a permutation matrix, we obtain the orthogonal NMF model. Orthogonal NMF is quite useful as it proves to be equivalent to  $K$-means clustering \cite{ONMF2006, SPL_AONMF}, where the columns of \mat{A} are the  cluster centroids and \mat{B} is the cluster indicator matrix. This representation is certainly unique as it meets the strongest form of the pure-source dominance condition. In summary, NMF and its various variants have been powerful and versatile tools in machine learning and signal processing.

\emph{Smooth Component Analysis (SmCA).} Smoothness is another important  intrinsic property
that provides  advantages, especially  for biomedical image and video processing.
In  SmCA,  we impose smoothness constraints not on the raw data
$\mat{X}$, but  rather on the hidden components, i.e., vectors of the factor matrix \mat{B}, 
and/or on basis vectors of the mixing matrix \mat{A}.
SmCA can be implemented by solving the following optimization problem \cite{Yokota-SmCA, YokotaZCY15}:
\begin{equation}
\min_{\mat{A},\mat{B}} 
\frob{\mat{X} - \mat{A}\trans{\mat{B}}}^2
+ \gamma_1 \frob[2,1]{\mat{L}_1 \mat{A}} +\gamma_2 \frob[2,1]{\mat{L}_2 \mat{B}},
\end{equation}
where {$\frob{\cdot}$ is the Frobenius norm defined in TABLE \ref{tab:notations}}, $L_{2,1}$ matrix norm is defined as $\frob[2,1]{\mat{Y}}=\sum_{t=1}^T\frob[2]{\mats[t]{y}}$ for $\mat{Y}\in\Real^{I\times T}$,  $\gamma_1,\gamma_2$ are nonnegative penalty parameters that 
control level of smoothness,
and  matrices $\mats[1]{L}, \mats[2]{L}$ represent difference operators, {typically,  the matrix with all its entries being zero, except 1 on its diagonal and $-1$ on its superdiagonal} \cite{NMF-book,YokotaZCY15, Yokota-SmCA}.

\begin{figure}[!t]
\centering
\includegraphics[width=.95\linewidth]{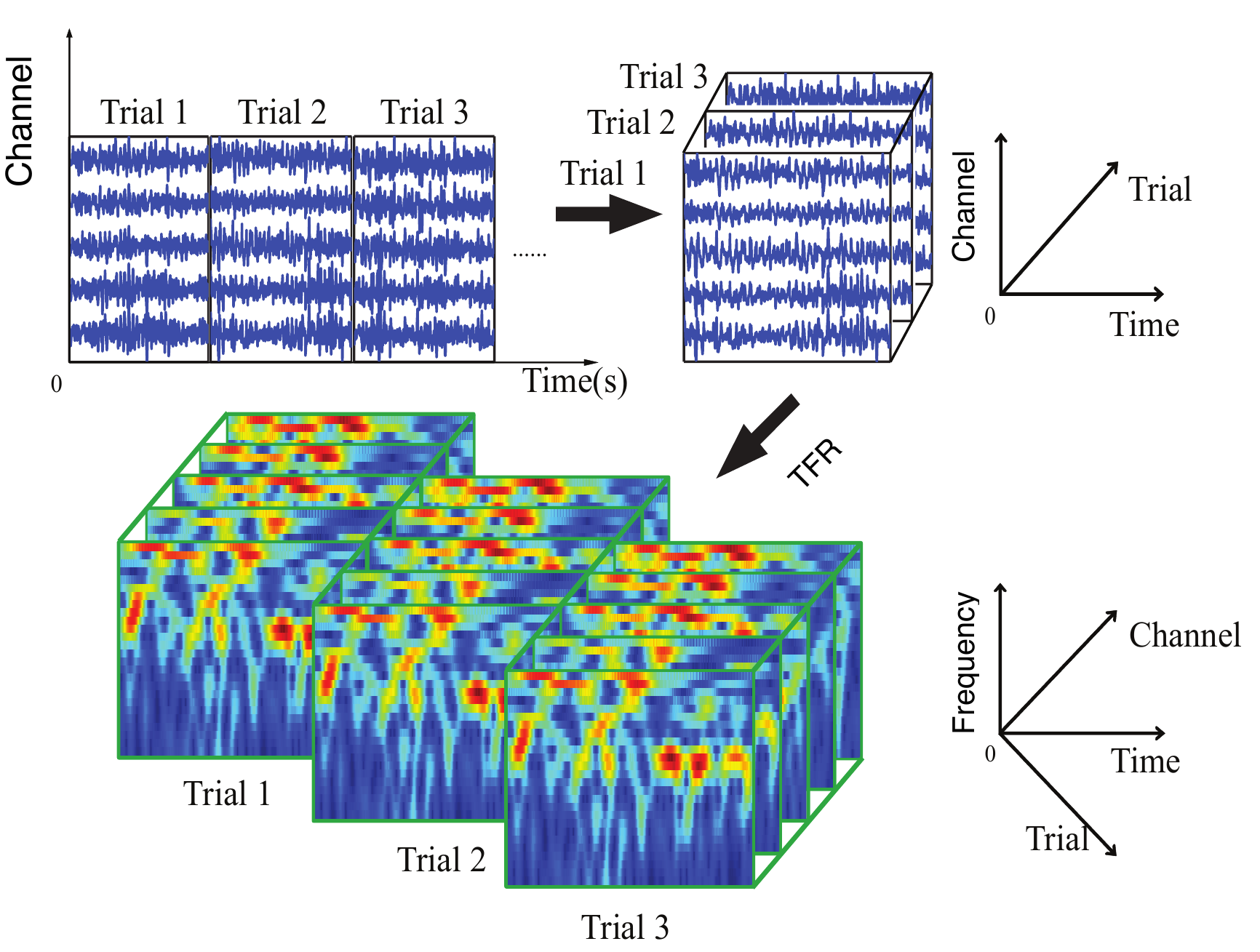}
\caption{Illustration of how to generate high-order tensor data for EEG data analysis, by incorporating the time-frequency representation (TFR) of each channel.}
\label{fig:tensorData}
\end{figure}

{After we build a proper objective function, BSS is finally achieved by solving an optimization problem. While many standard optimization approaches can be applied, some typical methods that are especially suitable for BSS are briefly introduced here. Taking into account that the mixing matrix in ICA is full column rank and hence has a certain underlying structure,  the concept of natural gradient  was proposed and proved to be the steepest direction \cite{NG_NIPS1996, SPL_NG1999}. On the other hand, for NMF and SmCA, alternating descent methods are quite useful since more than one parameter matrix need to be optimized. In these types of methods, each time the optimization is performed over a subset of parameters (typically, one factor matrix or even only one column of it) with the others fixed \cite{NMF-book,SPM_NMFNTD}.  The alternating direction method of multipliers (ADMM) \cite{ADMM2011},  which has been enjoying growing popularity recently and is well suited to large-scale optimization problems, also falls into this category. Basically, alternating descent is rather a principle than a concrete method and has been widely applied to solve large-scale optimization problems that arise in machine learning and related areas \cite{ADMM2011}.}

\section{Multiway Data Analysis}
So far the data we considered were represented as a single matrix, which has only two dimensions, or modes. In many applications the data can be naturally represented as high dimensional arrays, i.e., tensors. For example, in EEG, each trial can be formed by a matrix with two modes of channels (electrodes) and time, and multiple trials form a 3rd-order tensor of channel$\times$time$\times$trial. If we consider the time-frequency representation of  each channel, we  obtain a 4th-order tensor of channel$\times$time$\times$frequency$\times$trial, see \figurename \ref{fig:tensorData}. To analyze such type of data, we may simply unfold them to form a large matrix and then apply the variety aforementioned BSS methods. However, this often causes loss of internal structure of high-order data.  It is more desirable to take the tensor structure into account when we analyze such data.

\subsection{Basic Tensor Decomposition Models}
\begin{figure}[!t]
\centering{
\subfloat[Tucker model]{\includegraphics[width=.85\linewidth]{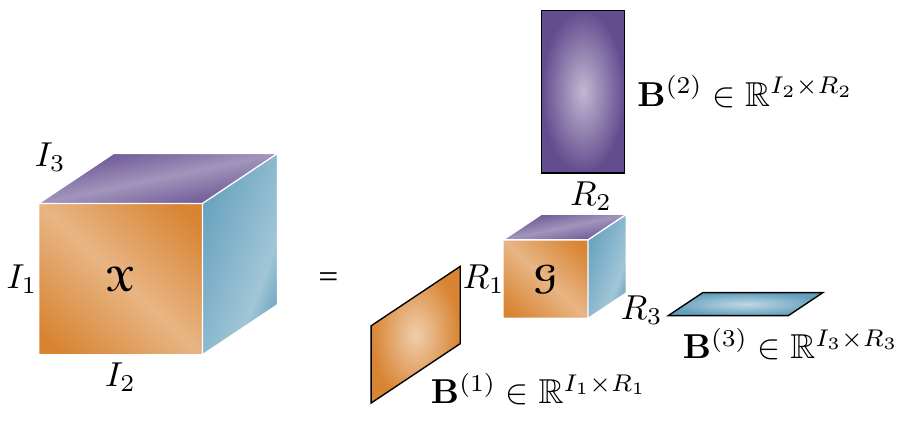}}\\
\subfloat[CP model]{\includegraphics[width=.85\linewidth]{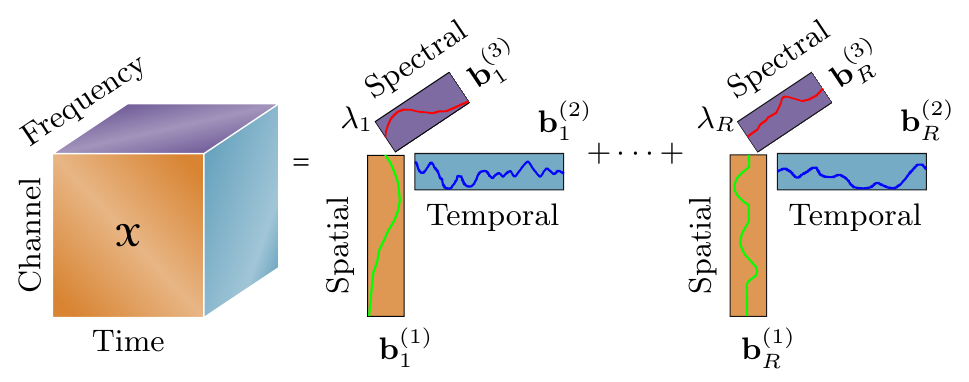}}
}
\caption{Illustration of the Tucker decomposition and the CP {decomposition} for 3rd-order tensors.}
\label{fig:TDModels}
\end{figure}

\emph{The Tucker Decomposition.}  Given an $N$th-order tensor $\tensor{X}\in\Real^{I_1\times I_2\cdots\times I_N}$, by Tucker decomposition, it can be represented as \cite{NMF-book,Kolda09tensordecompositions,SPM_Tensor}
\begin{equation}
\label{eq:Tucker}
\begin{split}
\tensor{X} &= \tensor{G}\ttmn[1]{B}\ttmn[2]{B}\cdots\ttmn[N]{B}\\
&\defeq\compactTucker{G}{B},
\end{split}
\end{equation}
where $\matn{B}=\begin{bmatrix}
\matn{b}_1 & \matn{b}_2 & \cdots & \matn{b}_{R_n}
\end{bmatrix}\in\Real^{I_n\times R_n}$ is the mode-$n$ factor component matrix consisting of latent components $\matn{b}_{r_n}$ as its columns,  $\tensor{G}\in\Real^{R_1\times R_2\cdots\times R_N}$ is the core tensor reflecting the connections (or links) between the components, and $(R_1,R_2,\ldots,R_N)$ with $R_n=\rank{\tenmat{{X}}}$ is called the multilinear rank of \tensor{{X}}.  \figurename \ref{fig:TDModels}(a) illustrates the Tucker decomposition of a 3rd-order tensor.

Unconstrained Tucker decomposition is generally not unique, because an equivalent representation can be obtained by replacing \matn{B} with $\matn{B}\mat{Q}$ and  \tensor{G} with $\tensor{G}\times_{n}\mat{Q}^{-1}$ in \eqref{eq:Tucker}, where \mat{Q} is any $R_n$-by-$R_n$ nonsingular matrix. In other words, \matn{B} estimates the range of \tenmat{X}. 
Hence, we often apply truncated SVD to mode-$n$ matricization \tenmat{X} for $n=1,2,\ldots,N$, in order to obtain orthogonal factor matrices \matn{B}. 
This method is well known as the high-order SVD (HOSVD) and has been widely applied to multidimensional data analysis \cite{HOSVD2000,DenoisingHOSVD2013, Phan2010TF, PNAS-HOSVD2007}.

\emph{The canonical polyadic (CP) decomposition}  (also known as PARAllel FACtor analysis (PARAFAC) or CANonical DECOMPosition (CANDECOMP)). If we restrict the core tensor \tensor{G} in   \eqref{eq:Tucker} to be diagonal with $R_1=R_2=\cdots=R_N=R$, we obtain the CP model that decomposes the target tensor into the sum of rank-one terms (see \figurename \ref{fig:TDModels}(b)):
\begin{equation}
\label{eq:Polyadic}
\begin{split}
\tensor{X} & =\sum\nolimits_r^R \;\lambda_r\;\matn[1]{b}_r\outerp\matn[2]{b}_r\outerp\cdots\outerp\matn[N]{b}_r \\
&\defeq\tenfactors{\mat{\Lambda};\matn[1]{B},\matn[2]{B},\ldots,\matn[N]{B}},
\end{split}
\end{equation}
where $\outerp$ denotes the outer product of vectors,
and $\mat{\Lambda}=\text{diag}(\lambda_r)$. The minimum number of $R$  is called the rank (or explicitly, CP-rank) of \tensor{X} \cite{Kolda09tensordecompositions}.

Using the mode-$n$ matricization operator, \eqref{eq:Tucker} and \eqref{eq:Polyadic} can be rewritten in matrix factorization forms as
\begin{equation}
\label{eq:TensorALS}
\trans{\tenmat{X}}=\matn{A}\trans{\matn{B}{}},\; n=1,2,\ldots,N,
\end{equation}
where {$\matn{A}={\left(\bigkkp\nolimits_{p\neq n}\matn[p]{B}\right)}\trans{\tenmat{G}}$} for the Tucker decomposition, or $\matn{A}=\left(\bigkrp\nolimits_{p\neq n}\matn[p]{B}\right)\mat{\Lambda}$ for the CP decomposition. Both models are multilinear models, and we often update the factor matrices  in an alternating fashion by solving $N$ least squares (LS) problems $\min\nolimits_{\matn{A}}\;\frob{\tenmat{X}-\matn{B}\trans{\matn{A}{}}}^2$, see \cite{Kolda09tensordecompositions,SPM_Tensor,SPM_NMFNTD, hunyadi2014block} for details.

Tensor decompositions have been widely applied to brain signal processing \cite{sanei2013eeg,FerdowsiAS13, shoker2005artifact,Cong2015EEG}.  In \cite{CPD_STFCA2009}, the CP decomposition was employed to extract time-frequency-space atoms to identify the different signal components of the EEG data, which renders tensor decomposition an efficient approach for processing EEG data. In \cite{TensorLocalize2014, CPDLocalize2007}, tensor-based EEG preprocessing was introduced to perform localization of brain sources based on EEG measurements, which can provide improved performance in the case of several simultaneously active brain regions and low signal-to-noise ratios. In \cite{ERPLAB}, a tensor toolbox for event-related potentials (ERP) analysis was developed, which includes multiway decomposition methods  and visualization of time-frequency transformed ERP data. The group analysis can be performed naturally by tensor representation of brain activities recorded from varying subjects, conditions or repeats. Based on several strategies to represent EEG in a tensor format, tensor decompositions, especially, nonnegative tensor decompositions, are applied to extract the dominant sources of brain activity  in an unsupervised manner, which can also be used for analysis of group differences between subjects.

As a natural extension of BSS, multiway BSS is an attractive and promising approach because {it allows components to be simultaneously extracted from different domains (i.e., modes) of tensor data}. For example, in EEG data analysis, via multiway BSS we can extract the spectral components, the temporal components, and the spatial components simultaneously, together with links between them, reflected by the core tensor.  As {essential} uniqueness is one key feature of BSS,  next we discuss unique tensor decomposition models and associated algorithms.

\subsection{Multiway Blind Source Separation}
As the Tucker decomposition is not unique in general sense, we often need to impose additional constraints on the factor matrices and the core tensor to obtain a unique and physically meaningful representation.  Comparing \eqref{eq:TensorALS} with \eqref{eq:MixingModel}, the columns of \tenmat{X} are just  linear mixtures of the columns of \matn{B}. This motivates us to extract \matn{B} by applying BSS algorithm to \trans{\tenmat{X}} directly, which has been discussed in \cite{MICA2005, MBSS2012} in terms of multilinear ICA and multiway BSS. Multiway BSS can be considered as an extension of the high-order SVD where {orthogonality constraints on factors are replaced with the other ones such as statistical independence, nonnegativity, sparsity, smoothness}, depending on \apriori knowledge about the components \cite{HOSVD2000,SPM_Tensor}.  
 As constrained Tucker decompositions, multiway BSS  is of particular interest in practice because it provides essentially unique  components that have specific statistical or deterministic properties (e.g., statistical independence, smoothness, sparseness and nonnegativity \cite{lraNTD_2015}). Particularly, different from 2D BSS, multiway BSS is able to accommodate all these diverse constraints or properties of factors in one Tucker model, allowing desirable interpretation of data and flexible feature extraction.

The CP decomposition can be interpreted as the Tucker model with additional structure constraints, and hence is essentially unique under the well-known sufficient condition \cite{Sidiropoulos2000}:
\begin{equation}
\label{eq:CPUniquess}
\sum_{n=1}^N\kappa_n\ge 2R+(N-1),
\end{equation}
where $\kappa_n$ is the Kruskal rank of \matn{B} {(see TABLE \ref{tab:notations})}, and $R$ is the rank of $\tensor{X}=\tenfactors{\mat{\Lambda};\matn[1]{B},\matn[2]{B},\ldots,\matn[N]{B}}$. More relaxed conditions can be found in \cite{CPDUniqueness2014}. This uniqueness analysis is purely based on multilinear algebra while neglecting the physical attributes of factors.

One of the most important  applications of the CP decomposition in BSS is   blind identification of the mixing matrix.
If we consider the ICA problem defined in \eqref{eq:traBSS}, the cumulant (can also  be any other high-order statistics) of ${\mat{x}}$ just forms a higher-order symmetric tensor
\begin{equation}
\label{eq:cumTensor}
\tensor{C}_{{\mat{x}}}=\sum_{r=1}^{R}{c}_r\;\mats[r]{a}\outerp\mats[r]{a}\outerp\cdots\outerp\mats[r]{a},
\end{equation}
where ${c}_r$ is the cumulant of ${s}_r(t)$, and $\mats[r]{a}$ is the $r$th column of \mat{A}. As a result, the mixing matrix $\mat{A}$ can be estimated via the CP decomposition of $\tensor{C}_{\mat{x}}$, and the identifiability is guaranteed by the uniqueness of the CP decomposition of $\tensor{C}_{\mat{x}}$. This approach is attractive because it is  able to estimate the  mixing matrix of even underdetermined mixing systems \cite{ComonBSS2010, Cardoso1998, FOBIUM, FOBIUM2007}.

If the sources have certain temporal structures, e.g., each source signal ${s}_r$ is either a deterministic ergodic sequence or a stationary  process \cite{Belouchrani1997, Choi2002}, the covariance matrices of ${\mat{x}}(t)$ satisfy
\begin{equation}
\label{eq:symJAD}
\mat{C}_{{\mat{x}}}(\tau)=\mat{A}\;\text{diag}(c_{{\mat{s}}}(\tau))\;\trans{\mat{A}},\quad \tau=0,1,2,\ldots,
\end{equation}
where $\tau$ is a time-lag, and $\text{diag}(c_{{\mat{s}}}(\tau))$ is a diagonal matrix whose diagonal entries are  $c_{{s}_r}(\tau)=\mathbb{E}[s(t+\tau)s(t)]$, see \cite{Belouchrani1997} for details. It can be seen that \eqref{eq:symJAD} is just the joint approximate diagonalization (JAD) of a set of symmetric matrices $\mat{C}_{{\mat{x}}}(\tau)$, $\tau=1,2,\ldots$, and has been extensively studied within the last two decades, see e.g., \cite{Belouchrani1997, Ziehe2004a, DeLathauwer:2006:Link, TNN_NJAD2009}. If we stack $\mat{C}_{{\mat{x}}}(\tau)$, we can obtain a 3rd-order tensor \tensor{C} satisfying
\begin{equation}
\label{eq:SOBICP}
\tensor{C}=\sum_{r=1}^T\mats[r]{a}\outerp\mats[r]{a}\outerp{c}_{{{s}}}(\boldsymbol{\tau}),
\end{equation}
where ${{{s}}}(\boldsymbol{\tau})$ is a vector of ${{{s}}}({\tau})$, $\forall\;\tau$. \eqref{eq:SOBICP} allows us to estimate \mat{A} by performing the CP decomposition of tensor \tensor{C}.
Again, the identifiability of \mat{A} is guaranteed by the uniqueness the CP decomposition of tensor \tensor{C}.

By comparing  \eqref{eq:TensorALS} with \eqref{eq:MixingModel},  we can also  incorporate ICA to CP decompositions.  A tensorial extension of probabilistic ICA (PICA), named tensor-PICA, was proposed in \cite{Beckmann2005} for multisubject fMRI analysis.  In tensor-PICA, assuming one component matrix, say \matn{B}, has mutually statistically independent columns (components),  a two-dimensional ICA criterion is used to update \matn{B} while the LS criterion is applied to update the other factor matrices, and all factor matrices are updated in an alternating manner. However, this method suffers from a divergence issue caused by the different objectives of the ICA step and the LS step \cite{MaartenDeVos2008}. In \cite{MaartenDeVos2008} a new method combining ICA and the CP decomposition was proposed, which overcomes the disadvantages of tensor-PICA but still suffers from high computational complexity. In \cite{SPL-SMBSS} a more efficient method was proposed, where one factor matrix \matn{B} was extracted by applying BSS to \tenmat{X}  (see \eqref{eq:TensorALS})  at first, then, by virtue of the special Khatri-Rao product structure of $\matn{A}=\bigkrp\nolimits_{p\neq n}\matn[p]{B}$, all the remaining factor matrices can be recovered by using a series of {SVDs} of rank-1 matrices. Particularly, once the matrix extracted by using BSS is of full column rank (which is almost always true), all the other factor matrices can be uniquely identified, even if the general algebraic uniqueness condition \eqref{eq:CPUniquess} for the CP decomposition {is} not satisfied, which somehow extends the  uniqueness analysis of the CP decomposition. In \cite{ERPUBSS2014} the robust  CP decomposition was also proposed to separate underdetermined mixtures of event-related sources that include quasi-periodic signals (e.g., electrocardiogram (ECG)), sources with synchronized trials (e.g.,  ERP), and amplitude-variant sources.

Note that under the nonnegativity constraints on the factors, the special Khatri-Rao product structure of $\matn{A}=\bigkrp_{p\neq n}\matn[p]{B}$ in \eqref{eq:TensorALS} often makes \matn{A} very sparse  \cite{SPM_NMFNTD}, even if \matn[p]{B} ($p\neq n$) have weak sparsity. 
The sparsity of \matn{B} allows us to  uniquely estimate \matn{B} and \matn{A} first by applying NMF to \tenmat{X}, and then recover all factor matrices \matn[p]{B} from \matn{A}, $p\neq n$, which is an efficient way to perform nonnegative CP decomposition (also known as nonnegative tensor factorization).
Nonnegativity is important for the CP decomposition because the optimal low-rank approximation under nonnegativity 
constraint always exists \cite{NTFComon2009} and is almost always unique \cite{uniNTF2015}, which is not true for  unconstrained CP decomposition. 
Hence, nonnegative CP decomposition is a very promising way to perform nonnegative multiway data analysis \cite{NMF-book,SPM_NMFNTD, nielsen2014non}.

\begin{figure*}[!t]
\centering
\includegraphics[width=.95\linewidth]{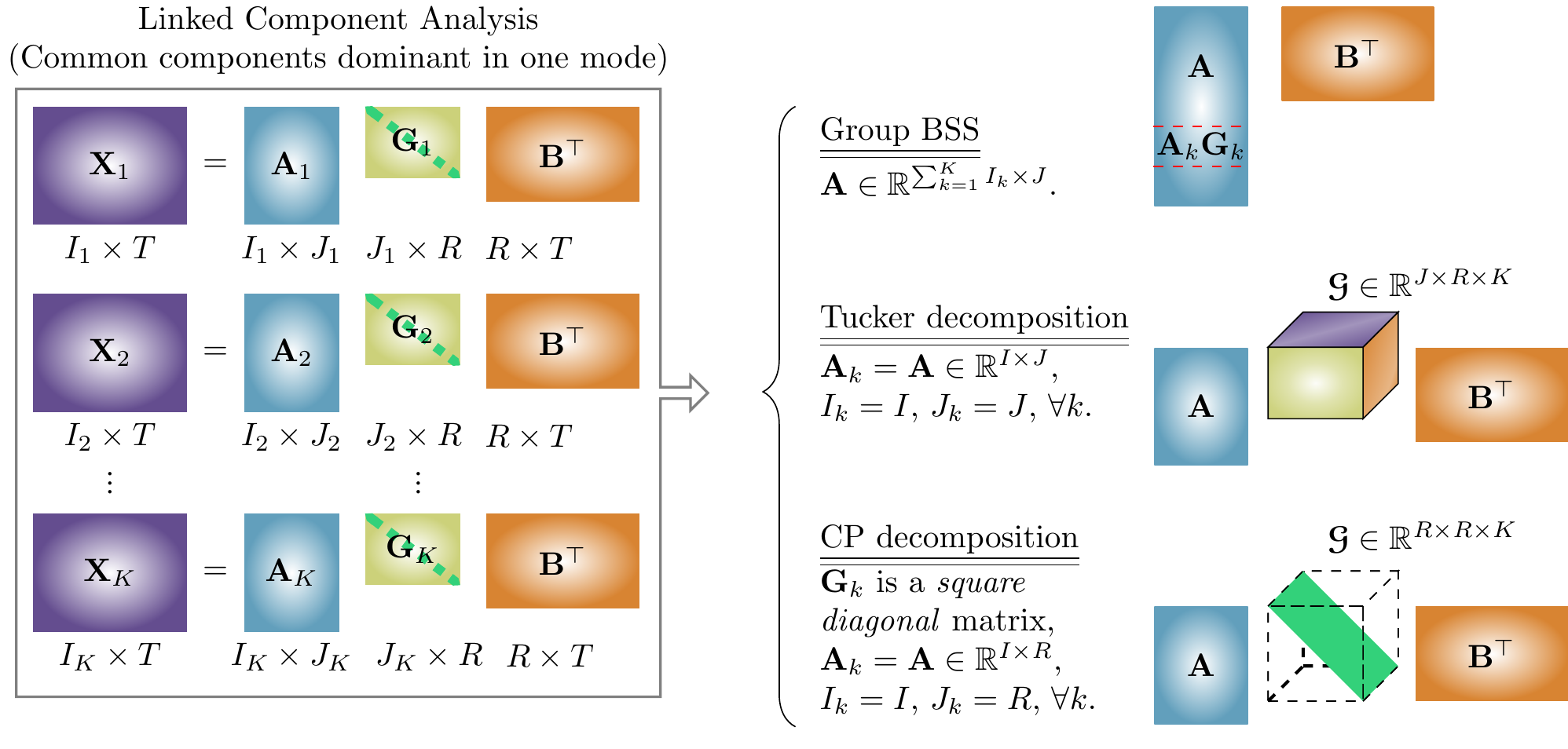}
\caption{Decomposition models interpreted as linked component analysis of multi-set data: $\mats[k]{X}=\mats[k]{A}\trans{\mats[k]{B}}=\mats[k]{A}\mats[k]{G}\trans{\mat{B}}$, where the latent components, i.e., the columns of \mat{B} are assumed to be exactly the same for all data blocks.}
\label{fig:groupAnalysis}
\end{figure*}

\renewcommand{\arraystretch}{1.5}
 \begin{table*}[!t]
\caption{Decomposition models for linked component analysis of multi-block data using common components}
\label{tab:linkedCA}
\centerline{
\colorbox{\figbkcolor}{
\begin{tabular}{ c | p{5in} }
\hline \hline
\multicolumn{1}{c|}{\bfseries Decomposition } & \multicolumn{1}{c}{\bfseries Linked component analysis} \\
\multicolumn{1}{c|}{\bfseries models} & \multicolumn{1}{c}{\bfseries ($k=1,2,\ldots,K$)} \\
 \hline
Concatenation-based models & $\trans{\mats[k]{X}}={\mat{B}} \trans{\mats[k]{A}}$ or $\mats[k]{X}=\mats[k]{A}\trans{\mat{B}}$, \\
joint ICA, group ICA, and variants& where \mat{B} contains the shared common components, see \eqref{eq:jointICA}. \\ \hline
\multirow{3}{*}{CP decomposition} & {$\mats[k]{X}=\mat{A}\; \mats[k]{\Sigma}\;\trans{\mat{B}}$}, where \mat{A} and \mat{B} contain the shared common components, see \eqref{eq:JADCP}. \\
  & \mats[k]{\Sigma} is a diagonal (scaling) matrix, \mat{A} can be an arbitrary matrix, or  has a {Khatri-Rao} product structure, \\
  & e.g., $\mat{A}=\mats[1]{M}\krp\mats[2]{M}$, which is equivalent to the CP decomposition of a higher-order tensor.\\  \hline
  \multirow{2}{*}{Tucker decomposition} & {$\mats[k]{X}=\mat{A}\,\mats[k]{G}\,\trans{\mat{B}}$}, \mat{A} and \mat{B} contain the shared common components, see \eqref{eq:PVD}. \\
  & \mat{A} is an arbitrary matrix, or has the Kronecker product structure, e.g., $\mat{A}=\mats[1]{M}\kkp\mats[2]{M}$, which is equivalent to the Tucker decomposition of a higher-order tensor.  \\
\hline \hline
\end{tabular}}
}
\end{table*}

\section{Linked Analysis of Multi-block Data}
Multi-block data  typically include multiple measurements of the same phenomenon under various experimental 
conditions. For example,
EEG signals or fMRI data 
of the human brain in response to a certain stimulus, but from different subjects and trials, can be grouped together and naturally linked as multi-block data. {Multi-block data may also refer to multi-modal data} where different types of pulse sequences 
can be used in a single scanning session to form  multimodal MRI datasets for each subject, or modalities such as fMRI and 
EEG might need to be considered jointly \cite{FerdowsiAS13}. All these 
possibilities pose the challenge to develop systematic approaches for fusing these various data types together to automatically find common patterns---if they exist---of related changes across multi-block data  \cite{acar2013understanding, AdaliIVASPM, LiangHNCC14}. Thus, linked component analysis is an emerging research topic with 
great potential in biomedical data analysis as it allows the extraction of not only shared common components across multi-block data but also individual components specific for each dataset.

A number of models have been introduced 
 to enable analysis of multi-block data by either defining hard links among matrices/tensors
 through shared factors, or by soft statistical links as in the case of independent vector 
 analysis (IVA) and canonical correlation analysis (CCA). To define these models, first
consider a set of data matrices $\mats[k]{X}\in\Real^{I_k\times T}$, $k=1,2,\ldots,K$.
Examples include brain imaging data where each $\mats[k]{X}$ is represented as 
voxel$\times$time points or voxel$\times$subject.  
These data blocks are naturally linked as they are collected across a single shared mode (time $T$), and can be represented as in \eqref{eq:MixingModel}  for all $K$ blocks (see \figurename \ref{fig:groupAnalysis})
\begin{equation}
\label{eq:linkedBSS}
\mats[k]{X}=\mats[k]{A}\trans{\mats[k]{B}},\; k=1,2,\ldots,K.
\end{equation}
In what follows, we first present models where the main assumption is the existence of common components
across all the datasets and is imposed through the model directly as hard constraints, then in Section 4.2, 
we consider the use of soft links that take advantage of statistical relationship, typically in the form of statistical dependence
as in IVA, CCA, and multiset CCA (MCCA), and finally introduce models that explicitly 
consider and model common as well as distinct components within each dataset.

\subsection{Models with common factors} 

To investigate shared common information across multi-block data, e.g., common spatial maps, a number of 
models have been proposed \cite{groupICA2008guo,GICA-rev,AdaliIVASPM}:

\emph{Concatenation-based ICA} models, as the name suggests, concatenate the data blocks either
in the vertical dimension as in Group ICA \cite{calhoun2012} or horizontally as in joint ICA \cite{jointICA}.
In a simple concatenation, the assumption is that all observations are spanned by a set 
of exactly the same components. For example, as illustrated in \figurename \ref{fig:groupAnalysis}, in 
\begin{equation}
\label{eq:jointICA}
\mats[k]{X}=\mats[k]{A}\trans{\mat{B}},\; k=1,2,\ldots, K,
\end{equation}
\mat{B} is shared by all data blocks whereas the mixing matrix $\mats[k]{A}\in\Real^{I_k\times R}$ is associated with \mats[k]{X} only.
By concatenating the observations vertically to form a tall matrix $\mat{X}=\trans{\begin{bmatrix}
\trans{\mats[1]{X}} & \trans{\mats[2]{X}} & \cdots &\trans{\mats[K]{X}}
\end{bmatrix}}\in\Real^{(\sum_kI_k)\times T}$, we have
\begin{equation}
\label{eq:ConcatICA}
 \mat{X}=\widetilde{\mat{A}}\trans{\mat{B}}
\end{equation}
where
$\widetilde{\mat{A}}=\trans{\begin{bmatrix}
\trans{\mats[1]{A}} & \trans{\mats[2]{A}} & \cdots & \trans{\mats[K]{A}}\end{bmatrix}}$.
Note that if horizontal concatenation is applied (assuming that $I_k=I$, $\forall k$ in \figurename \ref{fig:groupAnalysis}), we simply transpose \mat{X}, which leads to a model
that assumes a common mixing matrix---the matrix $\mat{A}$ in \eqref{eq:traBSS}---for the ICA
decomposition. This is the model adopted in  joint ICA  where we have 
${\mats[k]{X}}=\mat{A}\trans{\mats[k]{B}},\; k=1,2,\ldots, K,$ and the common 
(mixing) matrix ${\mat{A}}$ is used to fuse multi-modal data such as fMRI and EEG 
represented as {subjects by voxels and subjects by time points} respectively \cite{jointICA}.
In this application, the columns of the mixing matrix include subject 
covariations and when there are multiple groups, such as patients and healthy controls, 
biomarkers can be identified through a {\emph{t-test}} on these columns. Or, as explained
in \cite{GICA-rev}, multiple blocks  ${\mats[k]{X}}$ can be  
multi-task and multi-subject data of the same type and dimension and can be 
stacked either vertically or horizontally depending on the interpretation. An important step included in group ICA,
which has been implemented in the GIFT toolbox (available at \url{http://icatb.sourceforge.net}),
is double dimension reduction stages using PCA. If each ${\mats[k]{X}}$ represents a 
subject data such as fMRI (time points by voxels), then 
we first perform a subject level PCA, and after vertical concatenation of dimension-reduced
subject data, a second level PCA is applied at the group level to estimate a
common group subspace  \cite{calhoun2001,calhoun2012}. Individual subject maps
are then reconstructed using the group and subject-level PCA matrices thus
preserving most of the variability for individual subjects. Other implementations and 
applications of the group ICA model are also possible and discussed in \cite{GICA-rev}.

Compared with \eqref{eq:MixingModel}, the concatenated ICA model 
described in \eqref{eq:jointICA} is merely the partitioned version of standard BSS \eqref{eq:MixingModel}, which is a popular technique to solve the computational issue in the case where \mat{X} is extremely large in one dimension. 
In addition, this model allows for explicit modeling of common modes across
e.g., specific trials or subjects.  One major limitation of the joint ICA model is 
the strong assumption that all data blocks share exactly the same common independent components. 
The approach proposed in \cite{mCCA_Correa} uses CCA and MCCA \cite{kettenring} such that the
assumption of identical mixing matrix (hence subject covariations of \cite{jointICA}) is relaxed and
instead maximally correlated subject co-variations are estimated.

 \emph{Tensor ICA} stacks all the data blocks to create a 3rd-order tensor \tensor{X}, rather than 2D matrices obtained by concatenation. This technique has been used for multi-subject fMRI analysis based on the CP model, for tensors with dimensionality of voxel$\times$time$\times$subject \cite{Beckmann2005,LinkedICA,helwig2013critique}. In this method, data for each subject can be expressed as
 \begin{equation}
 \label{eq:JADCP}
 \begin{split}
\tensor{X}_{:,:,k}= \mats[k]{X} & =\mat{A}\;\text{diag}(\mats[k]{w})\;\trans{\mat{B}}\\
& = [\mat{A}\;\text{diag}(\mats[k]{w})]\;\trans{\mat{B}}.
\end{split}
 \end{equation}
 Note that the mixing matrix associated with \mats[k]{X} is $\mat{A}\text{diag}(\mats[k]{w})$, implying all blocks actually share the same mixing matrix but differently scaled by their individual weighting vectors \mats[k]{w}. This of course will lead to a more restrictive model in comparison with \eqref{eq:ConcatICA}, which can be clearly observed from the relationship
 \begin{equation}
 \label{eq:tensorICA}
  \trans{\tenmat[2]{X}}={(\mat{W}\krp\mat{A})}\trans{\mat{B}},
 \end{equation}
 where $\mat{W}=\begin{bmatrix}
\mats[1]{w} & \mats[2]{w} & \cdots & \mats[R]{w}
 \end{bmatrix}\in\Real^{K\times R}$, \tenmat[2]{X} is the mode-2 unfolding of the data tensor \tensor{X},  ${(\mat{W}\krp\mat{A})}$ serves as the mixing matrix and has a Khatri-Rao product structure,
and \mat{A} captures the latent spatial maps with additional independence constraints. Note that $\trans{\tenmat[2]{X}}$ is just the vertical concatenation of \mats[k]{X}. In other words, the major difference between the concatenation based ICA \eqref{eq:ConcatICA} and the tensor ICA \eqref{eq:tensorICA} is that:  In the concatenation based ICA, the mixing matrix is not constrained whereas in the tensor ICA it has a Khatri-Rao product structure. The authors of \cite{LinkedICA}  claimed that it can be a beneficial feature of the tensor ICA because it avoids unnecessary duplication of the spatial maps and can allow them to be inferred more accurately when the assumption holds.

It is worth noticing that \eqref{eq:JADCP} can be viewed as asymmetric joint diagonalization of matrices \mats[k]{X}, an extension of the symmetric JAD problem described in \eqref{eq:symJAD}, which shows that the CP decomposition in general is merely a technique that \emph{simultaneously diagonalizes} a set of  matrices (or more generally, tensors). Remembering the discussion in Section 2, 
SVD diagonalizes a single matrix with orthogonality constraints whereas the CP decomposition jointly diagonalizes multiple matrices but without additional constraints. Hence the CP decomposition is essentially a useful tool for linked component analysis of multi-block data that are of the same size and share common factors in each dimension. For example, in \cite{ShardCPD2013}, the CP decomposition was used for shared processing of perception and imagery of music, in order to determine if both the spatial location and temporal evolution {are} shared between tasks, identifying common networks and their timing signatures.

A natural question that may arise is  whether it is meaningful in practice to relax model \eqref{eq:JADCP} to
\begin{equation}
\label{eq:PVD}
\tensor{X}_{:,:,k}=\mats[k]{X}=\mat{A}\mats[k]{G}\trans{\mat{B}},
\end{equation}
where \mats[k]{G}, associated with \mats[k]{X}, is not necessarily diagonal any more. This interesting extension has been studied in \cite{PVD2011} in terms of population value decomposition (PVD) and has been applied to discover common features from thousands of similar images. This simple model allows to reduce {a very large number of
 images} to a manageable set of parameters and to make statistical inferences on samples of large-scale datasets. The authors demonstrated the potential of this method by applying it to analyze the sleep heart health study dataset. Moreover, the entries of matrix \mats[k]{G} can be used as predictors in a regression context, e.g., 
to predict the risk of Alzheimer's disease using fMRI \cite{PVD2011}.
If we stack \mats[k]{X} to form a 3rd-order tensor \tensor{X}, and similarly for \mats[k]{G},  by using the Tucker model, \eqref{eq:PVD} can be rewritten using tensor representation as (see \figurename \ref{fig:groupAnalysis})
\begin{equation}
\label{eq:PVDTucker}
\tensor{X}=\tensor{G}\times_1\mat{A}\times_2{\mat{B}},
\end{equation}
implying the multi-block data $\mats[k]{X}=\tensor{X}_{:,:,k}$ share common factors \mat{A} and \mat{B}, but with subject-specific information kept in $\mats[k]{G}=\tensor{G}_{:,:,k}$.

In summary, \eqref{eq:JADCP} and \eqref{eq:PVD} are special cases of the CP model and the Tucker model, respectively, which can be extended to higher-order tensors \cite{MRCPD} and applied to the scenarios where multi-block data are of the same size and share common components in all modes. Moreover, the concatenated ICA models
(group ICA and joint ICA), the CP decomposition, and the Tucker decomposition all perform linked analysis of multi-block data with the assumption that all samples are dominated by the common components, and with the exception of joint ICA that uses 
horizontal concatenation, all models assume the blocks to be of the same size. We summarize these
models in TABLE \ref{tab:linkedCA}.

\subsection{Models with statistical links: IVA, CCA, MCCA, PLS, and their tensorial extensions}

The methods discussed in Section 4 assume that the data blocks share exactly the same latent (independent) components, which could be too restrictive for some applications and lead to suboptimal solutions.
In contrast, IVA \cite{IVATSP2014,AdaliIVASPM}, CCA  \cite{hotelling} and its multiset extension MCCA \cite{kettenring}, as well as partial least squares (PLS) are more flexible methods, 
and provide advantages especially when not all modes are common across the blocks. 
It has been shown that MCCA can be used to perform BSS \cite{JBSS_MCCA}, fusion of EEG, fMRI and structural MRI data \cite{mCCA_Correa}, and for brain computer interface (BCI) \cite{IJNS_COBE}, and IVA for multi-subject fMRI analyis
\cite{AdaliIVASPM}, artifact rejection in concurrent EEG-fMRI data \cite{IVATBME2015} among others. 
In what follows, we 
first briefly review matrix decomposition using IVA, discuss how it generalizes 
CCA and MCCA---like the generalization of PCA using ICA---and then discuss the multilinear 
extensions for CCA and PLS \cite{KernelTensorsSPM}.

\begin{figure}[!t]
\centering
\includegraphics[width=.95\linewidth]{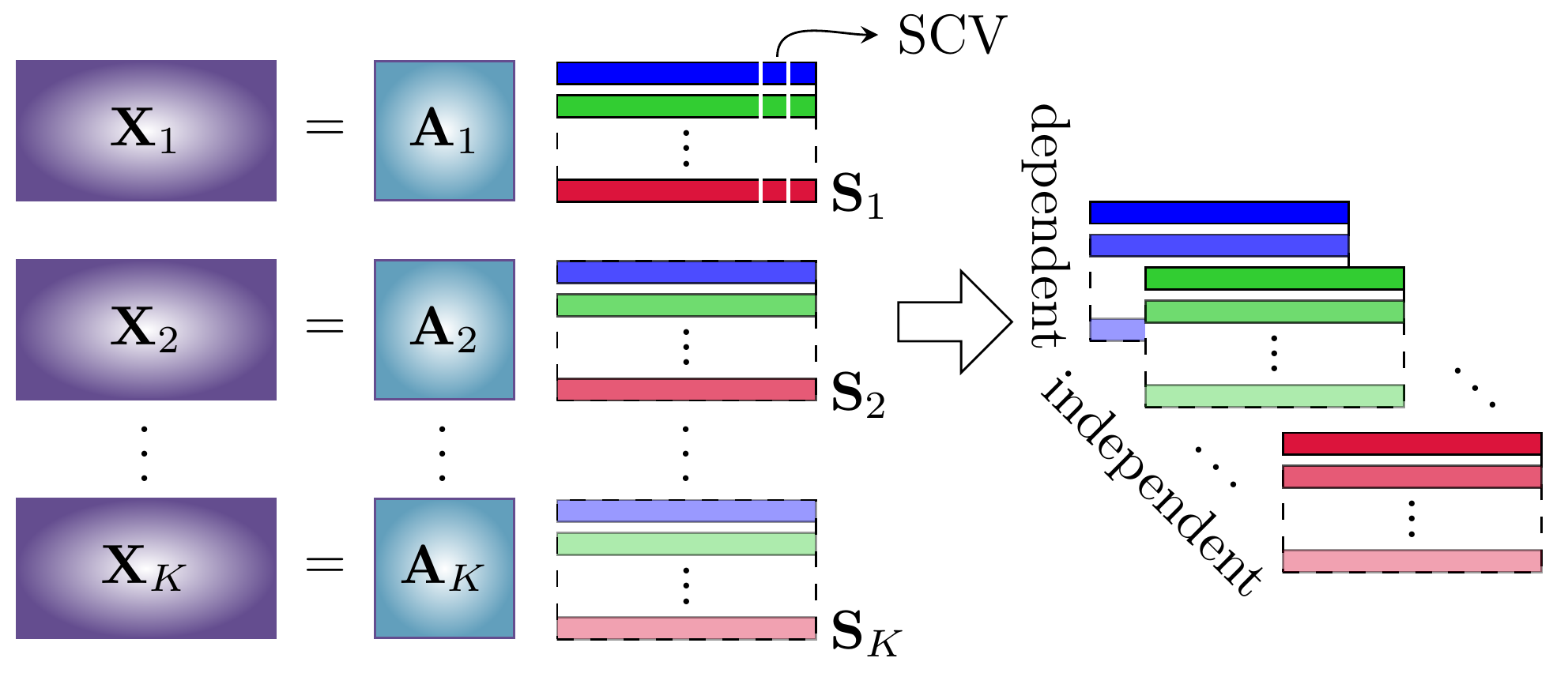}
\caption{Diagram of the IVA model.  }
\label{fig:iva}
\end{figure}

 \emph{Independent Vector Analysis (IVA)} can be formulated by considering the model in 
 \eqref{eq:traBSS} for $K$ datasets such that
 \begin{equation}
\label{eq:IVA}
{ \mats[k]{x}}(t)= \mats[k]{A}{\mats[k]{s}}(t),
\quad  k=1,2,\ldots, K.
\end{equation}
Hence, in this model, as opposed to those introduced in Section 4.1, all datasets are assumed to have their own unique 
mixing and source/component matrices. The link across the datasets is achieved in a ``soft manner'' 
by defining an optimization criterion
that fully takes advantage of the {statistical dependence that exists among the multiple datasets. }The key definition in the formulation of IVA is the source component vector (SCV) that is formed
by using the corresponding elements of the source random vectors ${\mats[k]{s}}(t)\in\Real^I$ such that an SCV is given by $\trans{[ s_{1,i}(t), s_{2,i}(t), \ldots, s_{K,i}(t)]}$ for $i=1,2,\ldots,I$, where the second subscript refers to the source index for each dataset in \eqref{eq:IVA}. 
{The IVA decomposition is achieved by minimizing the mutual information among the $I$, each $K$-dimensional SCVs, as opposed to sources in ICA  \cite{AdaliIVASPM, PIEEE_Adali_iva2015}. Hence, IVA achieves a decomposition of independent sources within each dataset while estimating SCVs that are dependent within its entries (see \figurename \ref{fig:iva}). 
It is this dependence that helps with the mitigation of permutation ambiguity for sources (modes) across the datasets, and hence enables IVA to identify sources that are independent and identically distributed Gaussians as long as they are correlated across the datasets. Complete identification conditions and bounds are given in \cite{IVATSP2014}. }
By properly aligning corresponding independent components from each dataset---whenever they are
related---IVA enables a convenient and useful way to perform multi-subject analysis as in fMRI
\cite{AdaliIVASPM} and allows one to exploit dependence for other uses such as in artifact rejection \cite{IVATBME2015}. {It has been shown that IVA is able to preserve subject variability compared to group ICA both by simulations \cite{michaelIVA2014, maicassp2013} and also using real data where there is significant {variations among the subject fMRI scans}, such as in patients with stroke \cite{laney2015capturing} and schizophrenia \cite{gopal2014}.}

When the SCV distribution is assumed to come from a multivariate Gaussian distribution, it can be shown that 
IVA cost function becomes equivalent to one of the five cost functions introduced for MCCA \cite{kettenring,AdaliIVASPM}. 
In \cite{kettenring}, maximization of correlation across elements of an SCV, and hence of multiple datasets, 
is achieved by making the covariance matrix of an SCV as ill-conditioned as possible 
through five cost-functions introduced by heuristic arguments, and each SCV is thus estimated in a deflationary manner. 
All five cost functions reduce to CCA when there are only two datasets. 
MCCA has been applied to both multi-modal data fusion and multi-subject fMRI analysis \cite{mCCA_Correa,JBSS_MCCA}
and hence IVA can be considered as a generalization of MCCA to the case where 
{higher-than second-order statistical relationship among the datasets are taken into account 
by selecting an appropriate multivariate density model for the SCV \cite{AdaliIVASPM}.}

\emph{Multilinear extensions of CCA and PLS.}
Suppose we have $K$ pairs of sample instances $\set{(\tensor{X}_k,\tensor{Y}_k),\; k=1,2,\ldots,K}$, with $\tensor{X}_k\in\Real^{I_1\times I_2\times\cdots\times I_N}$ and $\tensor{Y}_k\in\Real^{J_1\times J_2\times\cdots \times J_M}$, which can be concatenated to form two large tensors with additional sample dimension: $\tensor{X}\in\Real^{I_1\times\cdots\times I_N\times K}$ and $\tensor{Y}\in\Real^{J_1\times\cdots \times J_M\times K}$. We seek projection vectors such that the correlation between the multilinear projections of \tensor{X} and \tensor{Y} on them is maximized:
\begin{equation}
\label{eq:mlCCA}
\max\nolimits_{\set{\mat{w}_{\tensor{X}},\mat{w}_{\tensor{Y}}}}  \quad \frac{\trans{\mat{u}}\mat{v}}{\sqrt{\trans{\mat{u}}\mat{u}\trans{\mat{v}}\mat{v}}},
\end{equation}
where
\begin{equation}
\label{eq:CCAuv}
\mat{u}=\tenmat[N+1]{X}\mats[\tensor{X}]{w},\quad \mat{v}=\tenmat[M+1]{Y}\mats[\tensor{Y}]{w},
\end{equation}
 \mats[\tensor{X}]{w} and \mats[\tensor{Y}]{w} are just the vectorizations of rank-1 tensors $\tensor{W}_{\tensor{X}}$ and $\tensor{W}_{\tensor{Y}}$, respectively, and
 \begin{equation}
\label{eq:CCAWxy}
\begin{split}
\tensor{W}_{\tensor{X}} &=\matn[1]{w}_{\tensor{X}}\outerp\matn[2]{w}_{\tensor{X}}\outerp\cdots\outerp\matn[N]{w}_{\tensor{X}},\\ 
\tensor{W}_{\tensor{Y}} &=\matn[1]{w}_{\tensor{Y}}\outerp\matn[2]{w}_{\tensor{Y}}\outerp\cdots\outerp\matn[M]{w}_{\tensor{Y}}.
\end{split}
\end{equation}
The vectors \mat{u} and \mat{v} are called the first pair of canonical variables, reflecting the common information shared by the two groups \tensor{X} and \tensor{Y}. It turns out that \eqref{eq:mlCCA} can be easily solved via HOSVD, by applying the Lagrange multiplier method to \eqref{eq:mlCCA}. Then after deflation, by solving \eqref{eq:mlCCA} we seek the second pair  of canonical variables, and so on as
in MCCA. Note that the canonical variables can cross multiple modes of \tensor{X} and \tensor{Y}, as long as they are of the same size in these modes. Multilinear CCA has been extended to the kernel space and applied, for example, to behavior data classification \cite{KernelTensorsSPM}.

Basically, CCA enables extraction of components that are highly correlated 
and is widely applied for feature extraction. In contrast, PLS is a similar well-established framework with the aim to predict a set of dependent variables (responses) from a set of independent variables (predictors) through the extraction of a small number of latent variables. It solves the following optimization problem:
\begin{equation}
\label{eq:mlPLS}
\max\nolimits_{\set{\mat{w}_{\tensor{X}},\mat{w}_{\tensor{Y}}}}  \quad \trans{\mat{u}}\mat{v},
\end{equation}
where \mat{u} and \mat{v} are defined as in \eqref{eq:CCAuv}. Compared with \eqref{eq:mlCCA}, it can be seen that besides correlation, the energy (variance) of \mat{u} and \mat{v}  plays a key role in PLS. Again, \eqref{eq:mlPLS} can be solved by HOSVD. Once $\trans{\mat{u}}\mat{v}$ is maximized, we replace \mat{v}  in \eqref{eq:CCAuv} by \mat{u} that can be expressed by using \tensor{X} and $\mat{w}_{\tensor{X}}$. As such, a connection between \tensor{X} and \tensor{Y} is built, which allows us to predict $\tensor{Y}^{new}$ from a given new sample $\tensor{X}^{new}$. In practice, we often extract multiple pairs of vectors to improve the prediction accuracy, following a similar routine as for the multilinear CCA.

Two standard tensor decomposition frameworks (i.e., CP and Tucker) are both considered as the counterpart of SVD for tensorial data. The multilinear CCA and PLS described in (\ref{eq:mlCCA})--(\ref{eq:mlPLS}) are based on the CPD framework. Therefore, an alternative strategy is to apply Tucker decomposition in these methods, which results in a more flexible model in terms of fitting ability.  For instance, multiway PLS based on Tucker model, called HOPLS~\cite{TPAMI-HOPLS}, can be described as
\begin{equation}
\label{eq:HOPLS}
\max\nolimits_{\set{\tensor{G}_{\tensor{X}},\mat{w}_{\tensor{X}},\tensor{G}_{\tensor{Y}},\mat{w}_{\tensor{Y}}}}  \quad {\trans{\mat{u}}\mat{v}},
\end{equation}
where \mat{u} and \mat{v} are the same as in \eqref{eq:CCAuv}, but
 \mats[\tensor{X}]{w} and \mats[\tensor{Y}]{w} are the  vectorizations of $\tensor{W}_{\tensor{X}}$ and $\tensor{W}_{\tensor{Y}}$, respectively, and
\begin{equation}
\label{eq:PLSWxy}
\begin{split}
\tensor{W}_{\tensor{X}} &=\tensor{G}_{\tensor{X}}\times_1\matn[1]{W}_{\tensor{X}}\times_2\matn[2]{W}_{\tensor{X}}\times\cdots\times_N\matn[N]{W}_{\tensor{X}},  \\
\tensor{W}_{\tensor{Y}} &=\tensor{G}_{\tensor{Y}}\times_1\matn[1]{W}_{\tensor{Y}}\times_2\matn[2]{W}_{\tensor{Y}}\times\cdots\times_M\matn[M]{W}_{\tensor{Y}}. \\
\end{split}
\end{equation}
Hence, \eqref{eq:CCAWxy} can be considered as a special case of \eqref{eq:PLSWxy}; and HOPLS is a generalized framework for multiway PLS, which has shown the promising performance for decoding of movement trajectories of a monkey's hands based on 
ECoG signals~\cite{TPAMI-HOPLS,SPM_Tensor}. Similar strategy can be also extended to multilinear CCA.

\subsection{Common and Individual Feature Analysis (CIFA)}
\begin{figure}[!t]
\centering
\subfloat[CIFA for multi-block matrices]{
\includegraphics[width=.9\linewidth]{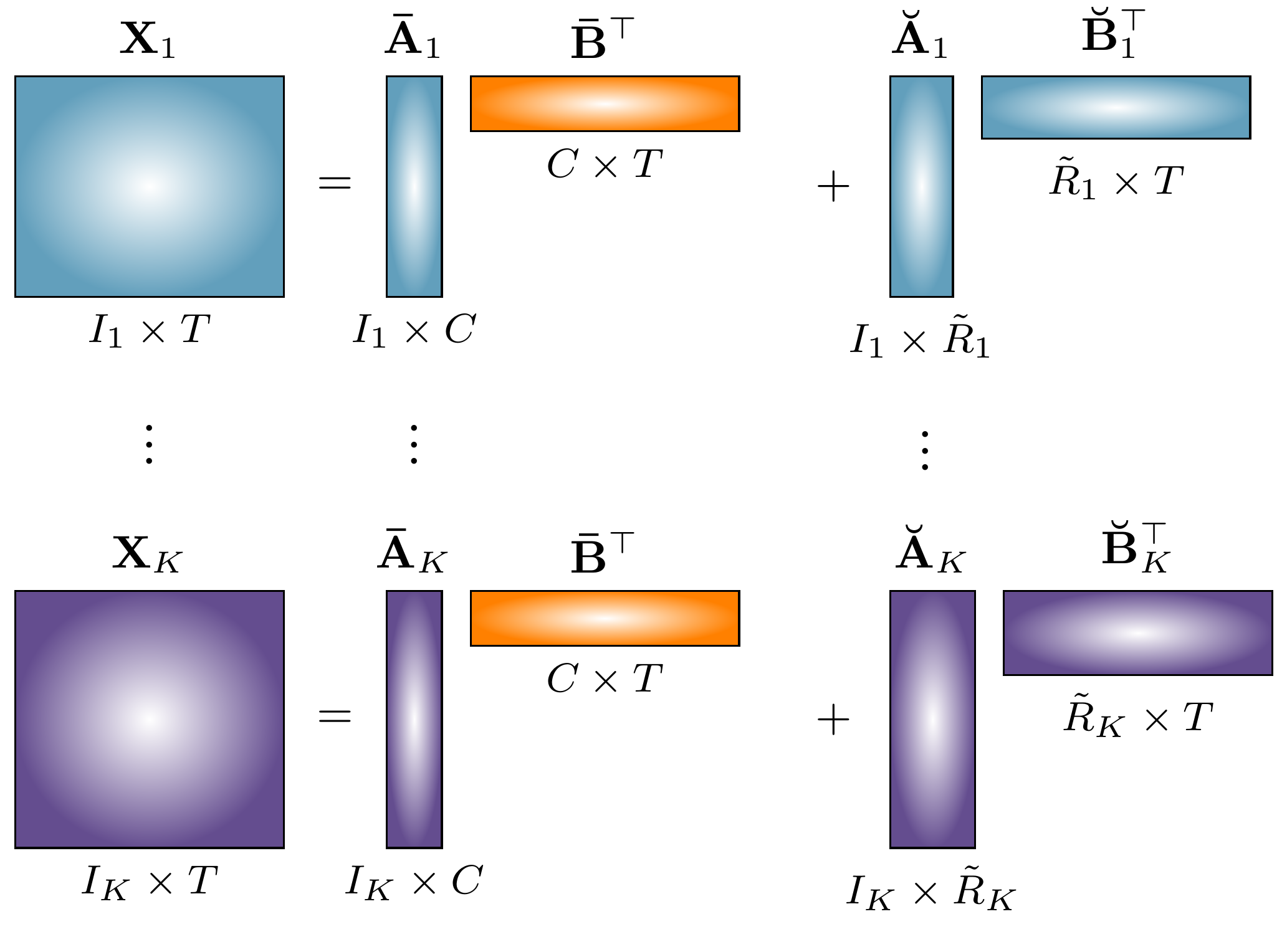}} \\
\subfloat[CIFA for multi-block tensors]{
\includegraphics[width=.9\linewidth]{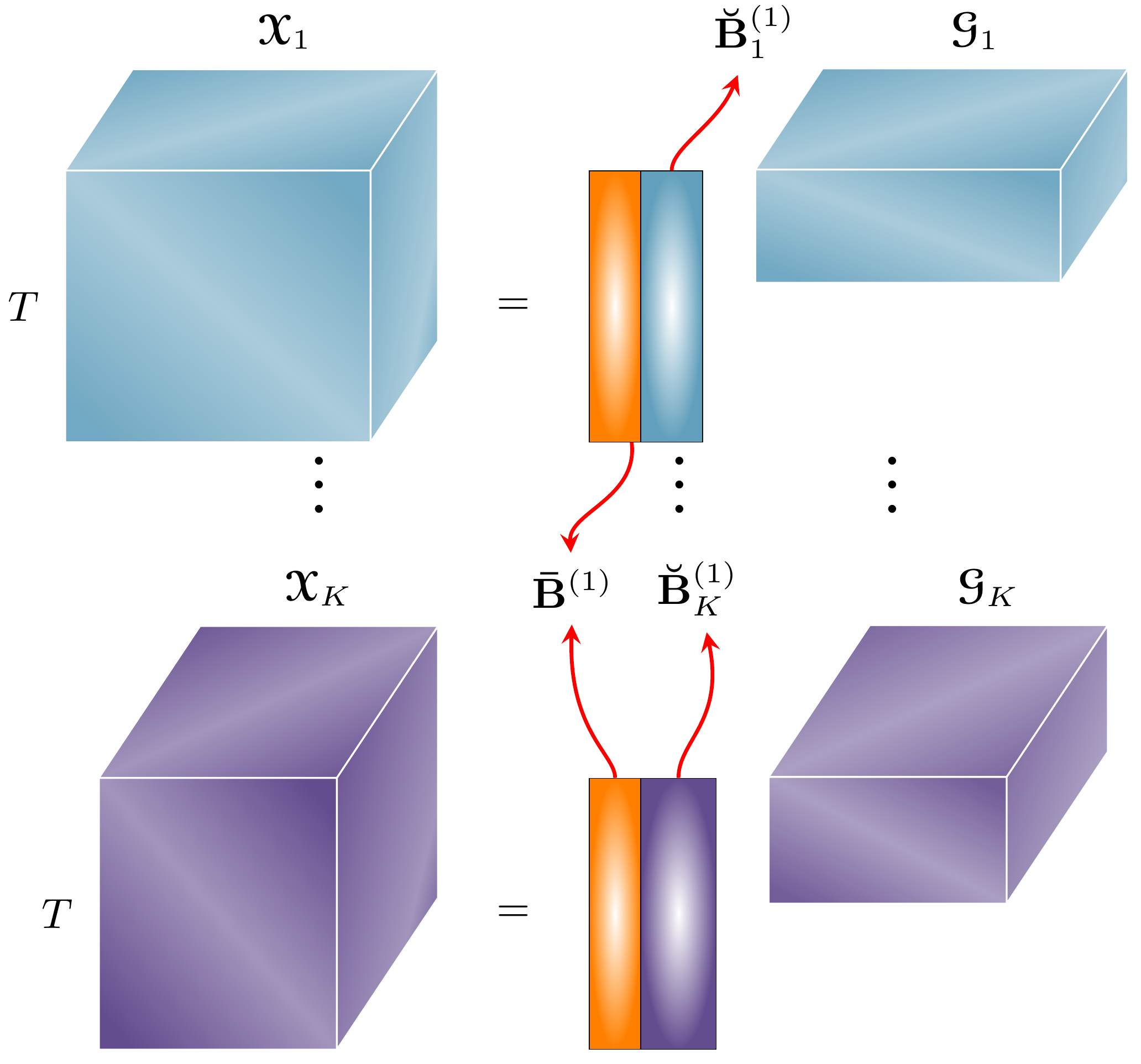}}
\caption{Illustration of common and individual feature extraction (CIFA). (a) The matrix case:  the matrix \mat{\bar{B}} denotes the common (or highly correlated) features shared by all data blocks while \mats[k]{\breve{B}} are the individual features specific for each individual block. (b) The tensor case: CIFA is performed with respect to mode-1, where the matrix $\matn[1]{\bar{B}}$ denotes the common features  while $\matn[1]{\breve{B}}_k$ are the individual features. The multilinear structure of $\tensor{G}_k$ makes the tensor model more general and flexible.}
\label{fig:CIFA}
\end{figure}

The models discussed in Section 4.1 assume that a certain set of components are exactly the same across all datasets, on the other hand, those discussed in Section 4.2 rely on statistical correlation (second and higher-order) 
to link the multiple datasets. Another approach is to explicitly model 
the common and distinct components, which is expected to be a better match to most physical systems, and hence 
lead to substantially improved performance. 
For example, if we analyze the human EEG signals in response to a certain stimulus, but from different subjects and for many trials, it is expected that  these data would share some common components due to the application of the same stimulus while also possessing components specific to each individual. To address such scenarios,  a more general linked matrix factorization model was recently studied in terms of a joint and individual variation explained (JIVE) \cite{JIVE2013}  and  common and individual feature analysis (CIFA) models \cite{JIVE2013,CIFA_ICASSP2015,CIFA-TNNLS}. In these methods, a set of  matrices $\mathcal{X}=\set{\mats[k]{X}\in\Real^{I_k\times T}: k=1,2,\cdots,K}$, share at least one common dimension $T$.  They can be, e.g., a set of multichannel EEG signals (channel$\times$time point) recorded for the same visual stimulus but from different subjects. These data matrices can be factorized in a linked way as follows (see \figurename \ref{fig:CIFA}(a))
\begin{equation}
  \label{eq:LinkedMF}
  \begin{split}
    \mats[k]{X} = \mats[k]{A}\trans{\mats[k]{B}}&=[\mats[k]{\bar{A}} \quad \mats[k]{\breve{A}}]\left[\begin{aligned} \mat{\bar{B}}\trans{} \\ \mats[k]{\breve{B}}\trans{} \end{aligned} \right]  \\
    = & \mats[k]{\bar{A}}\trans{\mat{\bar{B}}}+\mats[k]{\breve{A}}\trans{\mats[k]{\breve{B}}} \\
    \defeq& \mats[k]{\bar{X}}+\mats[k]{\breve{X}}, \quad \forall k,
  \end{split}
\end{equation}
where $\mat{\bar{B}}\in\Real^{T\times C}$, $\mats[k]{\breve{B}}\in\Real^{T\times {\tilde{R}_k}}$, $R_k=\tilde{R}_k+C$ is the number of latent components with $R_k< I_k$ and $C$ is the number of common components, $\mats[k]{\bar{A}}$ and $\mats[k]{\breve{A}}$ are the partitions of the mixing coefficients \mats[k]{A} corresponding to $\mat{\bar{B}}$ and $\mats[k]{\breve{B}}$. In this way, each data matrix \mats[k]{X} is represented through a combination of components from the common (shared) subspace $\mats[k]{\bar{X}}=\mats[k]{\bar{A}}\trans{\mat{\bar{B}}}$ and the individual (intrinsic) subspace $\mats[k]{\breve{X}}=\mats[k]{\breve{A}}\trans{\mats[k]{\breve{B}}}$. Basically, the common information carried by the common subspaces \mats[k]{\bar{X}} can be used to identify the whole group of coupled data, whereas the individual subspaces \mats[k]{\breve{X}} can be used to identify the data member \mats[k]{X} as it is  presented only in \mats[k]{X}. The objective is to estimate the matrix \mat{\bar{B}} representing common components shared by all data blocks and the matrices \mats[k]{\breve{B}} representing individual components associated with only the data matrix \mats[k]{X}.

\begin{figure*}[!ht]
\centering
\includegraphics[width=.85\linewidth]{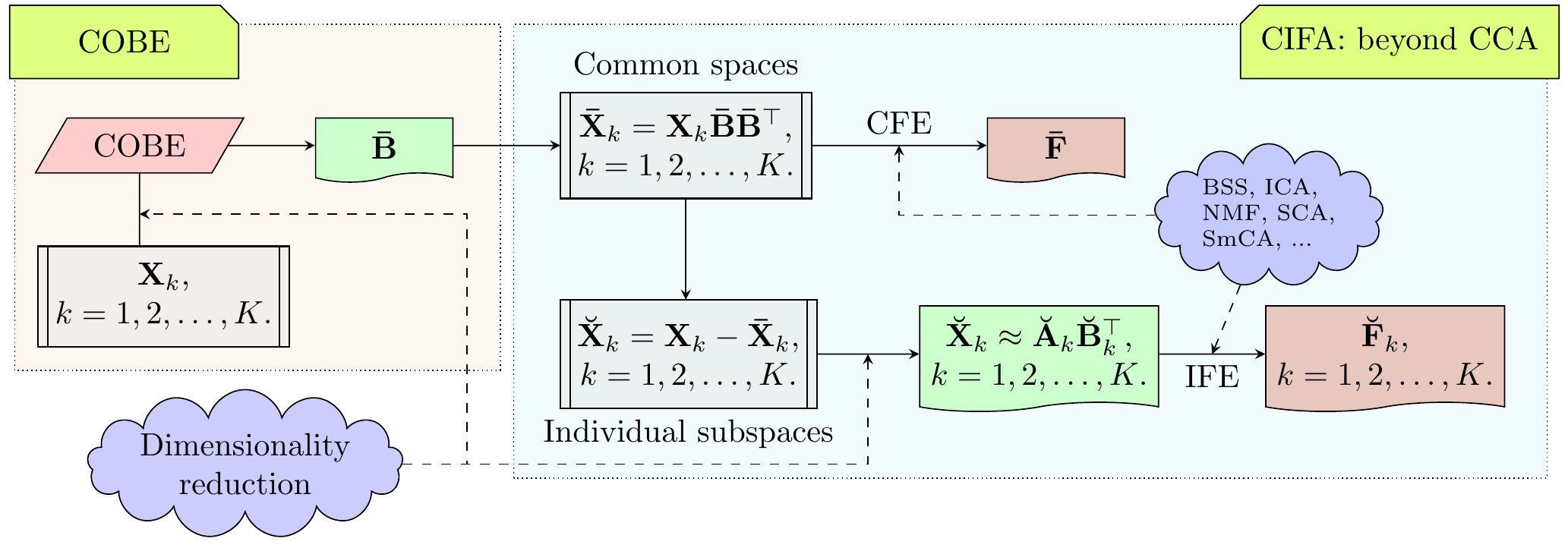}
\caption{{The paradigm of common and individual feature analysis (CIFA) for multi-block matrix data analysis. In CIFA, the common orthogonal basis extraction (COBE) approach \cite{CIFA-TNNLS} can be employed to separate the common and individual subspaces. Then, BSS can be respectively applied to the common and individual subspaces, 
to achieve  common feature (i.e., \mat{\bar{F}}) extraction and individual feature (i.e., $\mats[k]{\breve{F}}$, $k=1,2,\ldots,K$) extraction. }}
\label{fig:CIFAB}
\end{figure*}

The objective of both JIVE \cite{JIVE2013} and CIFA \cite{CIFA_ICASSP2015,CIFA-TNNLS} is to solve this challenging problem we describe above, with the subtle yet important difference: JIVE quantifies the amount of joint variation between data types whereas CIFA captures highly correlated components while ignoring their energy (variances), which allows  relatively weak common components to be detected and extracted. For this reason, JIVE can be viewed as the extension of PLS while CIFA is more closely related to CCA. CIFA advances this topic by introducing BSS methods to  the analysis of each common/individual subspace, motivated by the link between \eqref{eq:MixingModel}, \eqref{eq:jointICA}, i.e., by defining
\begin{equation}
\label{eq:linkedBSS}
\mats[k]{\bar{X}}=\mats[k]{\bar{A}}\trans{\mat{\bar{B}}},\quad
\mats[k]{\breve{X}}=\mats[k]{\breve{A}}\trans{\mats[k]{\breve{B}}}.
\end{equation}
This formulation allows us to  separate common and individual components that are physically easy to interpret. 
Hence CIFA enjoys significantly improved flexibility and has demonstrated numerous potential applications in clustering analysis and pattern recognition \cite{CIFA-TNNLS}. This paradigm is illustrated in \figurename \ref{fig:CIFAB}.

If each data block is  a high-order tensor rather than a matrix,  more sophisticated models are required to perform comprehensive linked analysis. For simplicity, at first we focus on the common and individual feature analysis with respect to only one mode, and then we briefly discuss its extension to multiple modes. To avoid cumbersome notations, sub- and super- scripts, we also assume that the tensors are of order 3, but please keep in mind that they can be of any order and may also have different sizes---as long as they share at least one common mode, from which common and individual components will be extracted. Consider a set of tensors \tensor{X}=\set{\tensor{X}_1,\tensor{X}_2,\ldots,\tensor{X}_K} that share a common mode, say, mode-1. Suppose $\tensor{X}_k\in\Real^{T\times I_{k,2}\times I_{k,3}}$, where  $\tensor{X}_k$ can also be a matrix with $I_{k,3}=1$, $\forall k$. Using the Tucker model, each tensor can be decomposed as
\begin{equation}
\label{eq:LinkedTucker}
\tensor{X}_k = \tensor{G}_k\times_1\matn[1]{B}_k\times_2\matn[2]{B}_k\times_3\matn[3]{B}_k,\;\forall k,
\end{equation}
where $\tensor{G}_k\in\Real^{R_{k,1}\times R_{k,2}\times R_{k,3}}$ is the core tensor of $\tensor{X}_k$, and $\matn{B}_k$, $n=1,2,3$, are the corresponding component matrices. Assuming that the mode-1 factor matrices $\matn[1]{B}_k$ share some common components such that $\matn[1]{B}_k=[\matn[1]{\bar{B}} \quad \matn[1]{\breve{B}}_k]$, we obtain the following linked Tucker decomposition model:
\begin{equation}
\label{eq:LinkedTuckerSplit}
\begin{split}
\tensor{X}_k  = &  \tensor{\bar{G}}_k\times_1\matn[1]{\bar{B}}\times_2\matn[2]{B}_k\times_3\matn[3]{B}_k \\
&   + \tensor{\breve{G}}_k\times_1\matn[1]{\breve{B}}_k\times_2\matn[2]{B}_k\times_3\matn[3]{B}_k \\
= &   \tensor{\bar{X}}_k+\tensor{\breve{X}}_k, \quad \forall k,
\end{split}
\end{equation}
where \matn[1]{\bar{B}} is shared by all data blocks whereas $\matn[1]{\breve{B}}_k$ is associated with $\tensor{X}_k$ only, $\tensor{\bar{G}}_k\in\Real^{C\times R_{k,2}\times R_{k,3}}$, $\tensor{\breve{G}}_k\in\Real^{(R_{k,1}-C)\times R_{k,2}\times R_{k,3}}$, and $C$ is the number of common components in mode-1.

For notational simplicity, we temporally denote the \emph{mode-1} matricization of $\tensor{X}$ by \tenmat[k,1]{X}, and similarly for $\tenmat[k,1]{G}$, $\tenmat[k,1]{\bar{G}}$ and $\tenmat[k,1]{\breve{G}}$; then \eqref{eq:LinkedTuckerSplit} can be rewritten in parallel with \eqref{eq:LinkedMF} as
\begin{equation}
\label{eq:LinkedTuckerMat}
\trans{\tenmat[k,1]{X}} = \matn[1]{A}_k\trans{\matn[1]{B}_k{}}= \matn[1]{\bar{A}}_k\trans{\matn[1]{\bar{B}}{}}+\matn[1]{\breve{A}}_k\trans{\matn[1]{\breve{B}}_k{}},
\end{equation}
where $k=1,2,\ldots,K$, and
\begin{equation}
\label{eq:LinkedTuckerM}
\begin{split}
 \matn[1]{\bar{A}}_k& =\left(\matn[3]{B}_k\kkp\matn[2]{B}_k\right)\trans{\tenmat[k,1]{\bar{G}}}, \\
 \matn[1]{\breve{A}}_k& =\left(\matn[3]{B}_k\kkp\matn[2]{B}_k\right)\trans{\tenmat[k,1]{\breve{G}}}.\\
\end{split}
\end{equation}
The key difference between \eqref{eq:LinkedTuckerMat} and \eqref{eq:LinkedMF} lies in the multilinear structures of  $\matn[1]{\bar{A}}_k$ and $\matn[1]{\breve{A}}_k$ shown in \eqref{eq:LinkedTuckerM}.

We call \eqref{eq:LinkedTuckerSplit} or \eqref{eq:LinkedTuckerMat} CIFA-Tucker as the constrained Tucker decomposition model is a key model of this linked component analysis. To solve this problem, we may apply the ALS method used in JIVE, but with additional multilinear structure constraints imposed on  $ \matn[1]{\bar{A}}_k$ and $ \matn[1]{\breve{A}}_k$. Alternatively, we can apply the two-step method proposed in  \cite{CIFA-TNNLS}: in the first step dimensionality reduction can be performed to obtain \eqref{eq:LinkedTuckerSplit}. Then, the rotational ambiguity associated with the Tucker model allows us to simultaneously rotate the columns of $\matn[1]{B}_k$ and conversely the rows of \tenmat[k,1]{G} to achieve  the separation of common and individual subspaces. Afterwards BSS can be applied to respective subspaces to extract unique and physically meaningful common and individual components with specific properties, such as sparsity, nonnegativity, statistical independence.

More generally, multi-block high-order tensor data may share more than one mode, each of which contains some common components.
To model such a  complex dataset, we can combine these modes to form a big mode that may contain some common components, by reshaping each data tensor $\tensor{X}_k$ simultaneously using the technique introduced in \cite{MRCPD}. As such, the matrix  \matn[1]{\bar{B}} in \eqref{eq:LinkedTuckerMat} is replaced by the Kronecker product of common components coming from different modes. In other words, we can impose additional structure constraints {upon} \matn[1]{\bar{B}} and $\matn[1]{\breve{B}}_k$. This problem can be solved using the ALS \cite{MRCPD}; implementation details are however omitted due to space limitation.

\renewcommand{\arraystretch}{1.5}
 \begin{table*}[!t]
\caption{Comparison of popular dimensionality reduction models (c.f. \cite{RGPCA2015}). $\mat{Y}\in\Real^{I\times T}$ is the data matrix of $T$ samples in $I$-dimensional space, $\mat{X}=\trans{\mat{AB}}$ is the extracted low-rank data,  the sparse matrix $\mat{E}$ denotes the outliers.  Laplacian matrix \mat{\Phi} characterizes a simple graph or a hypergraph between the samples of \mat{Y} \cite{RGPCA2015}, $L_{2,1}$  norm is defined as $\frob[2,1]{\mat{Y}}=\sum_{t=1}^T\frob[2]{\mats[t]{y}}$. Due to space limitation, most of their nonnegative variations are omitted here as they are straightforward, and their multiway extension can be found in \cite{qb_PAMI_CP,BayesCPD-TNNLS,zhao2015bayesian} and references therein.}
\label{tab:PCA}
\centerline{
\colorbox{\figbkcolor}{
\begin{tabular}{ c | p{2.25in} |p{2.2in} } 
\hline \hline
Model & Objective  & Constraints \\ \hline
PCA & $\min_{\mat{A},\;\mat{B}} \; \frob{\mat{Y}-\mat{A}\trans{\mat{B}}}^2$  & $\trans{\mat{A}}\mat{A}=\matI$ \\  \hline
RPCA \cite{rPCA2011}  & $\min_{\mat{X},\;\mat{E}} \; \frob[*]{\mat{X}}+\lambda\frob[1]{\mat{E}}$ & $\mat{Y}=\mat{X+E}$, $\mat{X}=\mat{A}\trans{\mat{B}}$\\ \hline
RPCA on graphs \cite{RGPCA2015} & $\min_{\mat{X},\;\mat{E}} \; \frob[*]{\mat{X}}+\lambda\;\frob[1]{\mat{E}}+\gamma\;\trace{\mat{X}\mat{\Phi}\trans{\mat{X}}}$ & $\mat{Y}=\mat{X+E}$, $\mat{X}=\mat{A}\trans{\mat{B}}$ \\  \hline
Graph Laplacian PCA (GLPCA) \cite{GLPCA2013} & $\min_{\mat{A},\;\mat{B}} \; \frob{\mat{Y}-\mat{A}\trans{\mat{B}}}^2+\gamma\;\trace{\trans{\mat{B}}\mat{\Phi}\mat{B}}$ & $\trans{\mat{B}}\mat{B}=\matI$ \\ \hline
Robust GLPCA \cite{GLPCA2013} &  $\min_{\mat{A},\;\mat{B}} \; \frob[2,1]{\mat{Y}-\mat{A}\trans{\mat{B}}}+\gamma\;\trace{\trans{\mat{B}}\mat{\Phi}\mat{B}}$ & $\trans{\mat{B}}\mat{B}=\matI$ \\  \hline
NMF \cite{Lee1999}& $\min_{\mat{A},\;\mat{B}} \; \frob{\mat{Y}-\mat{A}\trans{\mat{B}}}^2$  & \mat{A} and \mat{B} are nonnegative \\  \hline 
Orthogonal NMF  \cite{ONMF2006, SPL_AONMF} & $\min_{\mat{A},\;\mat{B}} \; \frob{\mat{Y}-\mat{A}\trans{\mat{B}}}^2$  & \mat{A} and \mat{B} are nonnegative, and $\trans{\mat{B}}\mat{B}=\matI$ \\  \hline
Sparse component analysis & $\min_{\mat{A},\;\mat{B}} \; \frob{\mat{Y}-\mat{A}\trans{\mat{B}}}^2+\lambda\frob[1]{\mat{B}}$  & Various constraints, e.g. nonnegativity  \\  \hline 
Smooth component analysis (SmCA) \cite{Yokota-SmCA} & 
$\min_{\mat{A},\mat{B}} \frob{\mat{X} - \mat{A}\trans{\mat{B}}}^2+ \gamma_1 \frob[2,1]{\mat{L}_1 \mat{A}} +\gamma_2 \frob[2,1]{\mat{L}_2 \mat{B}}$ 
& Various constraints, e.g., nonnegativity. $\mats[1]{L}, \mats[2]{L}$ represent difference operators. \\
\hline \hline
\end{tabular}}
}
\end{table*}

\section{Dimensionality Reduction for Incomplete  Data Corrupted by Outliers}
When we are ready to apply any specific method we have discussed to real-world problems, several practical issues arise.
These include questions such as {to}
how many latent components one should extract, and how to remove the noise and outliers from noisy and/or incomplete observation data.  Particularly,  multi-block high-order data may cause data deluge and pose new challenges for effective and efficient data analysis. A common approach to deal with these issues is dimensionality reduction, which has been a ubiquitous technique broadly applied for noise reduction, data compression, and feature extraction. Indeed,  most models and algorithms we discussed above rely on efficient and robust dimensionality reduction.

The key to achieve dimensionality reduction is the low-rank nature of data that is essentially related to the complexity of the  generative model. Estimating this rank, or equivalently, the number of latent components, is often a challenging task and referred to as {the} model selection problem.
As an example, in brain data imaging we never know how many latent sources that contribute essentially to the measured data; 
different model order selections may lead to quite different latent components and, consequently, different interpretations.

 A successful {{model order selection} largely relies on appropriate  assumption on the distribution of noises. Gaussian noise is usually assumed in many imaging systems \cite{HeZ2010-2}, which often results in a least squares problem and the use of PCA introduced in 
 Section 2. To deal with grossly corrupted observations, robust PCA (RPCA) was recently proposed and attracted  considerable attention. In RPCA, the noisy data \mat{Y} is modeled as the sum of a low rank term $\mat{X}=\trans{\mat{AB}}$ and a sparse term  \mat{E} of outliers, leading to the following optimization problem \cite{rPCA2011}:
\begin{equation}
\label{eq:RPCA}
\min_{\mat{X},\mat{E}} \; \rank{\mat{X}}+\gamma\frob[0]{\mat{E}},\quad s.t.\; \mat{Y}=\mat{X}+\mat{E}.
\end{equation}
However, \eqref{eq:RPCA} is a highly nonconvex optimization problem that is difficult to solve; a tractable surrogate of \eqref{eq:RPCA} is obtained by replacing $\rank{\mat{X}}$  by  the nuclear norm $\frob[*]{\mat{X}}=\sigma_i(\mat{X})$ and $\frob[0]{\mat{E}}$ by $\frob[1]{\mat{E}}$. By solving this surrogate problem, both low-rank approximation and the outliers can be estimated. Detailed discussions on this topic can be found in \cite{rPCA2011} and references therein.

A closely related problem, known as matrix completion (MC), also utilizes the low-rank feature of data and can be modeled as
\begin{equation}
\label{eq:MC}
\min_{\mat{X}}\; \rank{\mat{X}},\quad s.t.\; \mat{X}_{\Omega}=\mat{Y}_{\Omega},
\end{equation}
where $\Omega$ indicates location of available entries of \mat{Y}, $\mat{X}_{\Omega}$ is a matrix with  $[\mat{X}_{\Omega}]_{i,j}=x_{i,j}$ if $(i,j)\in\Omega$; otherwise 0. Apparently, if we know the location of outliers in \eqref{eq:RPCA}, problem \eqref{eq:RPCA} is equivalent to \eqref{eq:MC} because we can simply mark the items corrupted by outliers as missing values.  
We refer readers to \cite{MC_PIEEE} for a comprehensive overview, and here, 
focus on tensor completion. Tensor completion has many important applications: 1) it allows us to perform effective data analysis even if the data is incomplete or {corrupted by outliers};  however, these incomplete data are often simply discarded using traditional techniques; 2) it allows us to predict the missing items based on partially observed values, which is known as collaborative filtering and has been widely applied in recommender systems; 3) it promises exact recovery of signals from very limited samples. This feature makes the completion technique somewhat related to compressed sensing \cite{spm_cs2008}. However, completion mainly makes use of the low-rank property whereas the compressed sensing employs usually the sparsity of latent signals under a specific dictionary.

Extension of \eqref{eq:RPCA} and \eqref{eq:MC} to the tensor domain is rather straightforward if the Tucker model is used to model the data. For example,  $\rank{\mat{X}}$ can be simply replaced by the weighted sum of multilinear ranks of  tensor \tensor{X}, and then outlier detection and completion can be achieved by minimizing the multilinear rank. Efficient variations and algorithms can be found in \cite{TuckerC_TPAMI2013, RTucker:SIAM2014,infTucker2012}; a recent trend is to handle both outliers and missing values using a unified framework \cite{RTucker:SIAM2014}. Among these methods, the infTucker method \cite{infTucker2012} assumes that the multilinear rank is given and the other methods can learn the multilinear rank but  introduce new tuning parameters whose selection is often as difficult as the model order selection itself.

The rank minimization framework faces big challenges if the CP model is used to model the data, because even determining the CP-rank of a given tensor is NP-hard. If  the CP-rank is given, polyadic factorization with missing values has been developed by employing CP weighted optimization (CPWOPT) \cite{Acar-Morup11} and nonlinear least squares (CPNLS) \cite{CPNLS2013}. The  Bayesian CP developed in \cite{ICML2014-BayesCP} can infer the latent CP-rank, but it suffers from the weak point that  the missing data are handled {in} a heuristic way. Recently, a unified framework was proposed in \cite{qb_PAMI_CP, BayesCPD-TNNLS}, where the given data tensor \tensor{Y} (can be incomplete) is modeled as
\begin{equation}
\label{eq:fullBayesCP}
\tensor{Y}=\tensor{X}+\tensor{N}+\tensor{E},
\end{equation}
$\tensor{X}=\tenfactors{\mat{\Lambda};\matn[1]{A},\matn[2]{A},\ldots,\matn[N]{A}}$, tensor \tensor{N} is the Gaussian noise, \tensor{E} denotes the outliers. To infer the latent CP-rank $R$,   at the very beginning, $R$ will be assigned a sufficiently large value and then it shrinks gradually during the iteration process by enforcing column-wise sparsity of factor matrices. This method is {attractive due to} its  full Bayesian treatment which is able to simultaneously handle outliers, Gaussian noise, and incomplete data uniformly with automatic model order selection. Particularly, it is  free of any tuning parameters.

As a side comment, we would like to emphasize that outliers are not always treated as noise. For instance, in video surveillance analysis, outliers can be used to identify the activities that stand out from the background \cite{BayesCPD-TNNLS}. Therefore, an alternative interpretation of robust tensor factorization is to decompose an incomplete tensor as  a low-rank tensor, capturing global information and predicting missing values, and a sparse tensor, capturing local information. Hence, the power of the methods we introduced in this section are not limited {to} optional pre-processing procedures. Instead, many of them can be applied independently  for different types of applications. {Dimensionality reduction itself is a very active research topic,  many models and approaches have been proposed for different application purposes. While readers may find comprehensive reviews in \cite{spm_dr2011}, we  summarized some of them for quick reference in TABLE \ref{tab:PCA}.}

It can be seen that the low-rank nature of data is the foundation of the above models and approaches. This low-rank nature allows a very compact representation of large-scale structured raw data, upon which highly scalable and efficient algorithms can be developed \cite{TSP-lraNMF, SPM_NMFNTD}. To meet the ongoing challenge created by big data analytics, new data models and optimization algorithms are required. In this regard,  tensor networks were recently proposed to achieve the so-called super-compression (see \cite{ciaTN2014, ciaBigData2014}  and references therein).
Randomization and sparsification are emerging  techniques that provide favorable scalability for dimensionality reduction. In the first category, the given huge data tensor can be compressed via either random projections \cite{RandomizedLowRank_Michael_2011, siam_probLowRank} or random sampling \cite{CURTensorCaiafaC2010}. In the second category,  the dense data tensor is converted to a sparse one via an element-wise sparsification algorithm \cite{tensorSparsification2015}.  In \cite{spm2014_ConvOpti}, the authors argued that  first-order methods, randomization, and parallel and distributed computation are three key pillars and offer surprising scalability benefits for big data analysis. Readers are referred to  \cite{ciaTN2014,spm2014_ConvOpti} and references therein for this ongoing research topic.}

\section{Illustrative Examples}
\subsection{Applications of CIFA in SSVEP-based BCI}
Steady-state visual evoked potentials (SSVEP) are  periodic neural responses elicited at the same frequency as a visual stimulus when a subject is excited at a specific frequency by the visual stimulus. An SSVEP-based BCI is designed to detect the target stimulus by recognizing the dominant frequency from SSVEP data, and then to translate it into a pre-assigned computer command. 
Although real SSVEPs  present relatively stable spectra as long as the recording time is sufficiently long, they are often contaminated by background sources of noise. Thus, efficient and accurate recognition of SSVEP frequencies is essential to achieve high performance in SSVEP-based BCIs. As one of the most widely adopted methods, CCA  identifies each SSVEP target frequency by locating the maximum correlation between EEG recordings and a set of artificial references, each of which corresponds to a candidate frequency \cite{SSVEPCCA2007}. 

\begin{figure}
  \centering
  \includegraphics[width=\linewidth]{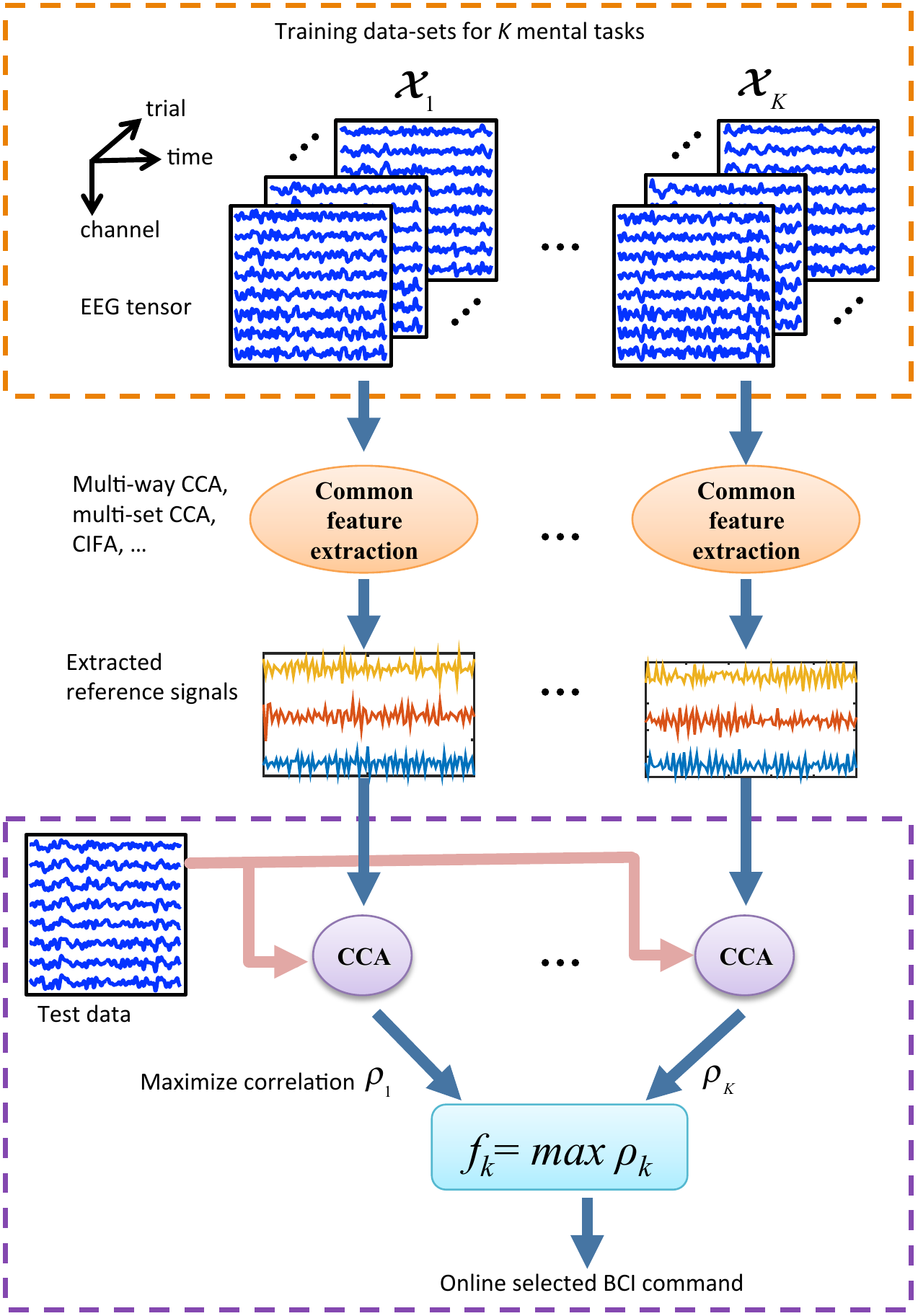}\\
  \caption{Illustration of  SSVEP recognition based on multiway or multiset learning models. Consider multiple tensors (channel $\times$ time $\times$ trial) $\mathcal{X}_1,\ldots,\mathcal{X}_K$ constructed by the EEG data corresponding to   $K$ distinct target frequencies of $f_1,\ldots,f_K$, respectively. The  common features  extracted by using MWCCA, MCCA or CIFA are used as references for SSVEP recognition. The SSVEP target {frequency of new test data for a single  trial} is then recognized according to the maximum of correlation coefficients $\rho_1,\ldots,\rho_K$ between the test data and the optimized references.}
  \label{fig:SSVEPCFE}
\end{figure}

While CCA gained popularity and provides reasonably good performance in SSVEP frequency recognition, it suffers from two major limitations: it  ignores the multiway structure of EEG data and particularly, uses artificial references typically constructed by simple sine and cosine waves rather than real SSVEP features. Recently,  by effectively exploiting the multiway feature of EEG data, a multiway extension of CCA (MWCCA) has been developed to improve the performance of SSVEP recognition \cite{L1MWCCA}. The MWCCA approach collaboratively maximizes correlation between multiple dimensions of EEG tensor data and the pre-constructed reference sine-cosine waves at specific stimulus frequencies, thereby leading to improved performance against the standard CCA. 
On the other hand, a group of EEG data trials recorded at the same stimulus frequency should share some common features reflecting this frequency information. Such common features extracted from EEG training data are supposed to bear real SSVEP characteristics and hence are more qualified as references for SSVEP recognition (see \figurename \ref{fig:SSVEPCFE}). With this intuition, MCCA was proposed to optimize the SSVEP references \cite{IJNS_COBE}, where the common features extracted from multiple training sets of EEG data recorded at the same stimulus frequency were chosen as the references.  A solution that does not use any reference signal but makes use of the statistical dependence of  SSVEP among multiple electrodes has been very recently proposed using IVA and has noted to provide quite promising performance \cite{SSVEPciss2015} provided that the time window is sufficiently long.

\begin{figure}
  \centering
  \includegraphics[width=0.9\columnwidth]{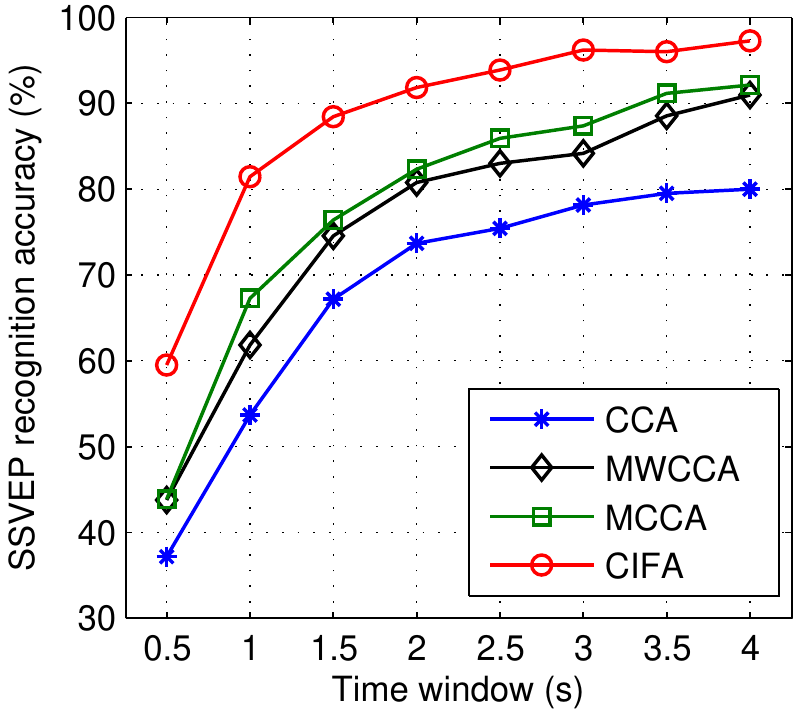}\\
  \caption{SSVEP recognition accuracies averaged on subjects, derived by CCA, MWCCA, MCCA and CIFA respectively, {over a} time window from 0.5 s to 4 s (0.5 s interval). }
  \label{fig:SSVEP}
\end{figure}

Also, we investigated the application of CIFA to learn real SSVEP features in order to further improve the accuracy of SSVEP recognition. In this experiment, we validated effectiveness of  CIFA  in SSVEP recognition using the EEG dataset recorded from ten subjects. During the EEG recordings, four red squares were presented on the screen as stimuli flickered with different frequencies. Each subject completed 20 runs. In each run, each stimulus was cued as the target followed by flickering of the four stimuli at four different frequencies: 6, 8, 9 and 10 Hz. Subjects were asked to focus their attention on the target for {1--4s} after each cue. A total of 80 trials data were acquired from each subject. More details about the experiment settings can be found in  \cite{L1MWCCA}. Three state-of-the-art methods, i.e., CCA, MWCCA and MCCA were compared with CIFA. \figurename \ref{fig:SSVEP} depicts the averaged recognition accuracies evaluated by leave-one-run-out cross-validation, from which we see that  CIFA achieved much higher average accuracy than the others  for all time windows from 0.5s to 4s.

\subsection{MRI denoising}
\begin{figure}
  \centering
  \includegraphics[width=0.7\columnwidth]{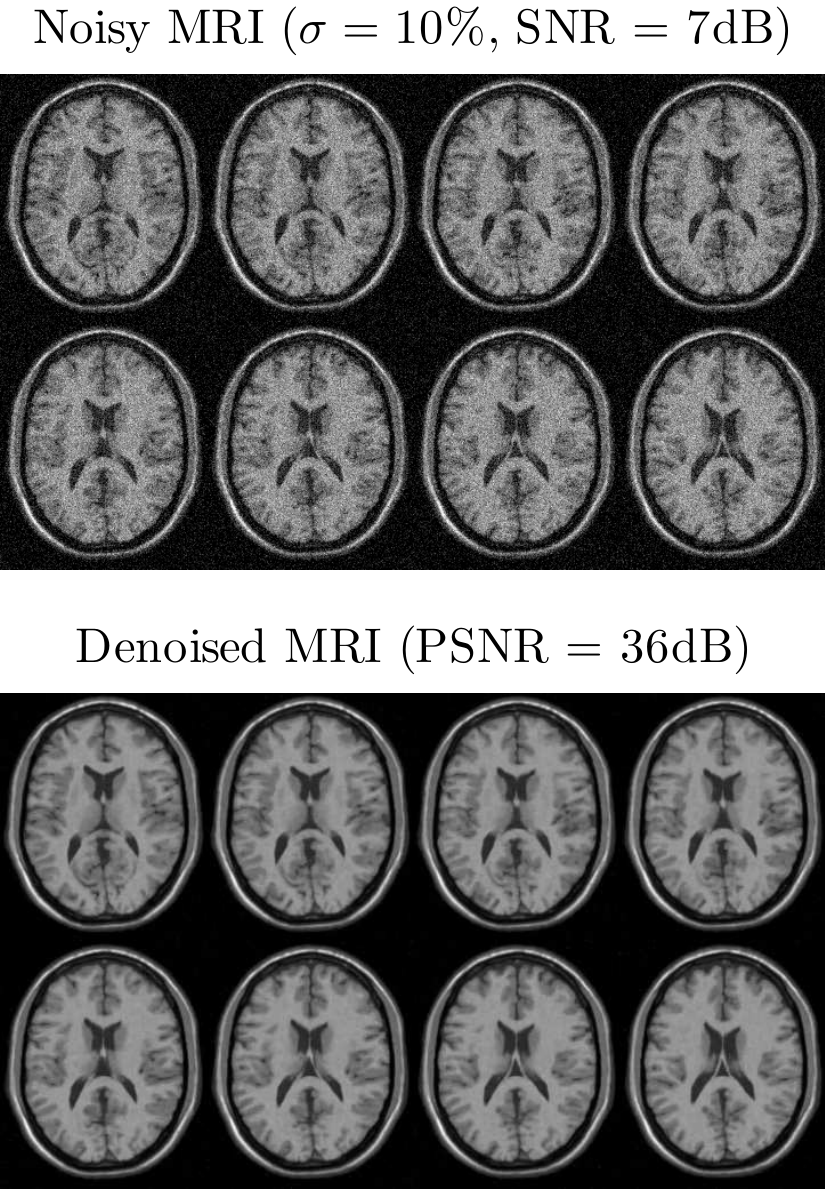}\\
  \caption{MRI tensor is of size $181\times 217\times 165$. The cube size is $4\times 4 \times 4$ and the number of similar cubes is 30.}
  \label{fig:MRIdenoising}
\end{figure}

MRI is a non-invasive high resolution multi-parameter imaging modality that is widely used in current clinical diagnosis and scientific research because of its ability to reveal the 3D internal structures of objects. However, MRI may suffer from serious 
effects such as random noise, specifically when high-resolution and high-speed is required. The denoising of MRI is thus important to improve the inspection quality and reliability of quantitative image analysis. Nonlocal filters by exploiting similarity and sparseness among cubes have been demonstrated to be effective in denoising MRI. Recently, HOSVD has been applied to image denoising by exploiting redundancy in the 3D stack of similar patches~\cite{DenoisingHOSVD2013}. In addition, the augmented HOSVD method has been investigated and applied to MRI volume data~\cite{zhang2015denoising}. One common limitation of existing denoising methods is that the noise level is assumed to be known in advance. Therefore, it is appealing to perform MRI denoising automatically without needing tuning of the parameters.  To achieve this, we apply a recently proposed  Bayesian CP factorization (BCPF)~\cite{qb_PAMI_CP} method for MRI denoising.  The procedure is described briefly as follows. Given an $I\times I\times I$ reference cube in the noisy MRI data, $K$ similar cubes (including the reference cube) are found and stacked into a 4D array of size $I\times I\times I\times K$, which is thus filtered by BCPF. Finally, the denoised MRI is obtained by aggregating multiple estimations at each location. Since BCPF can estimate CP-rank efficiently and the noise variance automatically from given data, it can be easily applied to filtering 4D array instead of HOSVD. For illustration, we use the public MRI data\footnote{Dataset is available from \url{http://brainweb.bic.mni.mcgill.ca/brainweb/}} and consider Gaussian noise {whose} standard derivation is 10\% of brightest tissue. Fig.~\ref{fig:MRIdenoising} shows the results of MRI denoising  with an 
impressive performance of PSNR$=36$dB.

\subsection{MRI completion}
\begin{figure}
  \centering
  \includegraphics[width=0.9\columnwidth]{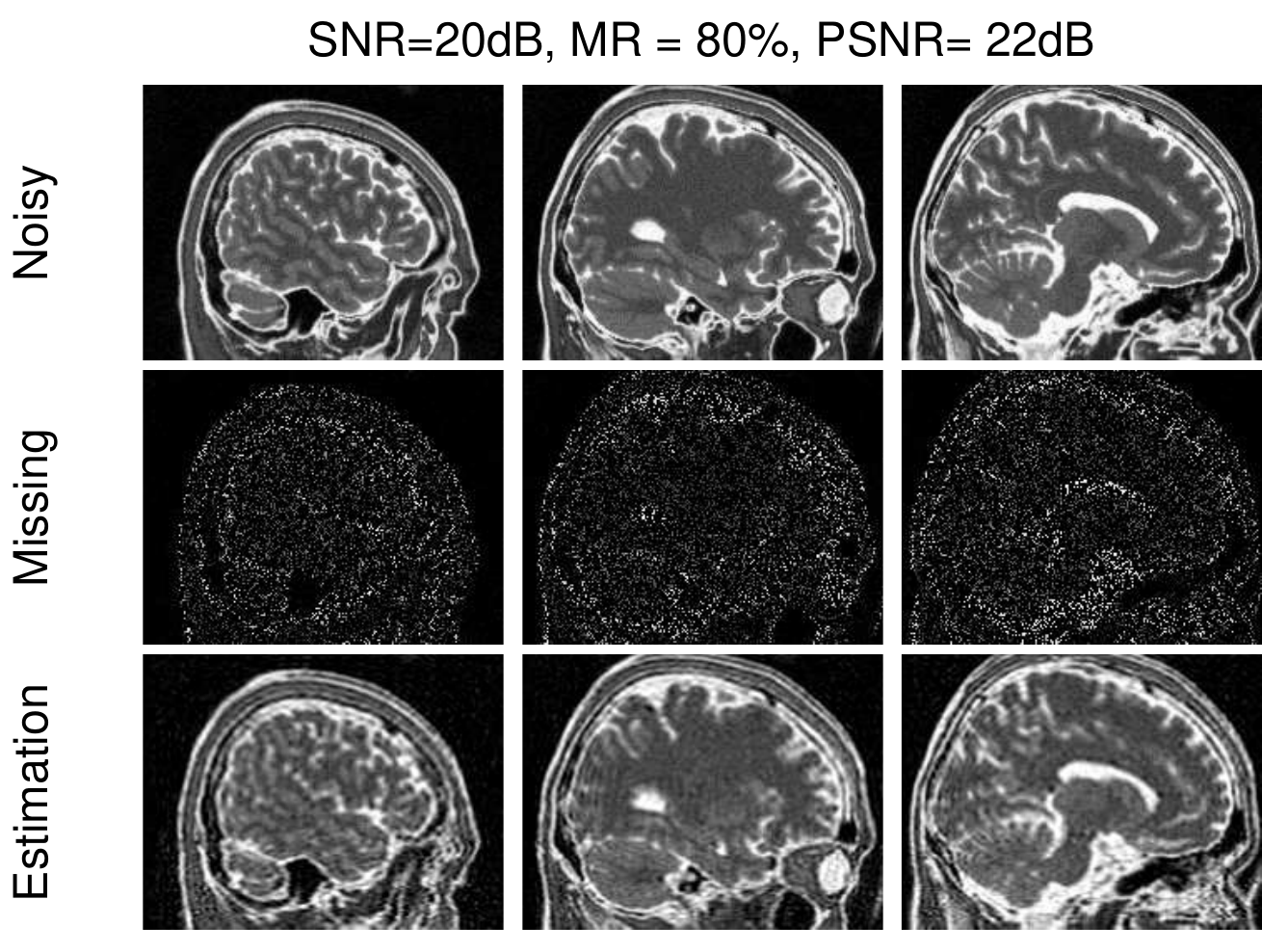}\\
  \caption{Tensor completion for noisy MRI data. The standard derivation of Gaussian noise was 3\% of brightest tissue. MRI tensor is of size $181\times 217\times 165$. }
  \label{fig:MRIcompletion80}
\end{figure}

Missing data are a common problem in MRI studies due in large part to susceptibility artifact and image acquisition parameters that result in incomplete brain coverage and spatial variation in acquired images, across subjects. In addition, MRI completion can be also used to improve spatial and temporal resolution by reconstructing the whole image from very few measurements. In this experiment, we illustrate MRI completion by applying  the Bayesian tensor completion (BTC) method~\cite{zhao2015bayesian}, particularly to improve the generalization performance when original data 
do not posses a global low-rank structure. Several recently proposed tensor completion methods including {{the high accuracy low rank tensor completion (HaLRTC)
method~\cite{TuckerC_TPAMI2013}, the generalized higher-order orthogonal iteration (gHOOI) method~\cite{liu2014generalized}, and the one based on weighted Tucker (WTucker) decomposition~\cite{filipovic2013tucker} }}
were also used for comparisons. For HaLRTC {and WTucker}, the tuning parameters were  selected to yield the best possible performance. BTC methods as well as gHOOI can automatically adapt the tensor rank to data.  As shown in Table~\ref{tab:MRIresults}, BTC achieves the best performance under the conditions studied here. When missing ratio is low (e.g., 50\%), HaLRTC outperforms gHOOI, while gHOOI outperforms HaLRTC when {the} missing ratio is high (e.g., 80\%).  The visual quality of reconstructed MRI obtained by  BTC is shown in Fig.~\ref{fig:MRIcompletion80}.

\begin{table}
\renewcommand{\arraystretch}{1.1}
\caption{\small The performance of MRI completion evaluated by PSNR and RRSE under missing ratios of 50\% and 80\%. }
\label{tab:MRIresults}
\centering
\begin{tabular}{ c |c | c | c | c }
\hline
  & \multicolumn{2}{c|}{50\% missing}  & \multicolumn{2}{c}{80\% missing} \\
  \cline{2-5}
Methods  & PSNR  &  RRSE & PSNR  &  RRSE  \\
 \hline
BTC &  {\bf 26.40}  & {\bf 0.12}  & {\bf22.33} &  {\bf 0.19}    \\
WTucker & 23.18  &  0.18 & 21.53  & 0.21   \\
gHOOI & 21.57  &  0.21 & 19.77  & 0.26   \\
HaLRTC & 23.12  &  0.18 & 17.06  & 0.36   \\
\hline
\end{tabular}
\end{table}

\section{Discussion, Conclusions and Future Perspectives}
Multi-set or multi-block coupled and inhomogeneous data are becoming ubiquitous across the sciences and engineering, especially in 
the biomedical field. Large scale medical/biological data using modalities such as EEG, fMRI, ECoG
critically require scalable algorithms to efficiently process massive datasets within reasonable amount of  time.  
To handle these issues, tensors, multi-dimensional generalizations of matrices, are attractive and promising tools 
for the representation and processing of such data.
 
In this paper, we have studied analysis of such multi-relational or multi-block data using simultaneous (coupled) constrained matrix and tensor factorizations that allow one to extract common and individual components and/or  to establish links between them. 
We critically reviewed several useful matrix and tensor decomposition models for linked multiway analysis 
(or multilinear blind source separation).
 
For large-scale multi-relational data analysis, tensor decompositions is one emerging technology. Current applications in many areas of science and technology, such as computational neuroscience, bioinformatics, and pattern/image recognition, generate massive amounts of data with multiple aspects and high dimensionality, and a number of successful examples have 
demonstrated that {tensor} decompositions with low-rank approximations can provide often useful and meaningful approximate representations
for such large scale applications \cite{ciaTN2014, papalexakis2014turbo}. 
Tensor and matrix decompositions, in particular using the linked models introduced here, 
allow us to discover meaningful hidden structures of complex brain data and perform generalizations by capturing multi-linear and multi-aspect relationships. 
To serve this purpose, flexible CIFA models were introduced to simultaneously model  common features shared by all data sources and individual features that are specific to each data source.  
In BSS, subspace definitions for {a} joint BSS framework as recently introduced \cite{Lahat2014,Silva2014}
can be used to define such common and distinct subspaces within IVA and the concentrated (hence linked) ICA models like group and joint ICA.
For all those methods, a number of challenges exist for very large scale data, feature extraction/selection, classification, clustering and anomaly detection.
A recent paper addresses the selection of model order when the sample size is relatively small by specifically considering 
common and distinct subspaces as in CIFA \cite{Song2015,CIFA-TNNLS}.

Another closely related approach  for big data not discussed in this paper is tensor networks. Complex interactions and operations  between tensors can be visualized by tensor network diagrams in which  tensors are represented graphically by  nodes or shapes \cite{ciaTN2014, ciaBigData2014}.
In such representations, each outgoing edge (line)  emerging from a node represents a mode (a way, a dimension, indices).  
Tensor  network diagrams could be potentially  very helpful not only in visualizing tensor
decompositions but also for expressing complex  mathematical (multilinear) operations of contractions of tensors \cite{ciaTN2014, ciaBigData2014}.
Tensor networks are  connected to   quantum physics, quantum chemistry and
quantum information, which studies the ways to possibly build a quantum computer and to program it  \cite{Orus2013}. In addition, tensor networks provide graphically illustrative large distributed networks and enable one to perform complex tensor operations (i.e., tensor contractions and reshaping)  in an intuitive way and  without the need to use explicit  mathematical representation.

To summarize, the benefits of linked multiway  (tensor) analysis methods along with matrix-based approaches  
for massive biomedical data include:
\begin{itemize}
 

\item  A framework to incorporate  different type of diversity or various constraints in different modes or  factors,
and naturally extending the standard component analysis and BSS methods to large-scale joint BSS as well as multiway  data analyses;
 
     \item Possibility to handle noisy and  missing  data naturally by using powerful low-rank tensor approximations   and by exploiting robustness and stability of tensor
         decomposition algorithms;

    \item ``Super''  compression of large-scale multidimensional data via tensorization
     and  decomposition of a high-order tensor into factor matrices and/or core tensors with low-rank and low-order  \cite{SPM_Tensor, ciaTN2014};
 
     \item Ability to perform all mathematical operations in feasible tensor network formats, with the ability to improve
     the scalability of algorithms \cite{ciaTN2014}.
 \end{itemize}

Hence, 
linked component analysis is a very promising area of research with substantial potentials in biomedical applications.

\bibliographystyle{IEEEtran}

\end{document}